\documentclass{emulateapj}
\newcommand{\erg}{\mbox{$\rm\,erg$}}
\newcommand{\kev}{\mbox{$\rm\,keV$}}

\newcommand{\ev}{\mbox{$\rm\,eV$}}

\newcommand{\s}{\mbox{$\rm\,s$}}
\newcommand{\ks}{\mbox{$\rm\,ks$}}

\newcommand{\kpc}{\mbox{$\rm\,kpc$}}

\newcommand{\beq}{\begin{equation}}
\newcommand{\eeq}{\end{equation}}

\shorttitle{The {\it Chandra} Gas Gallery of Normal Ellipticals}
\shortauthors{Diehl \& Statler}

\usepackage{natbib}

\begin{document}

\title{The Hot Interstellar Medium of Normal Elliptical
  Galaxies. I.\linebreak A {\it Chandra} Gas Gallery and Comparison of X-
ray and Optical Morphology}

\author{Steven Diehl\altaffilmark{1,2} and Thomas S. Statler\altaffilmark{1}}
\altaffiltext{1}{Astrophysical Institute, Department of Physics and Astronomy,
251B Clippinger Research Laboratories, Ohio University, Athens, OH 
45701, USA}
\altaffiltext{2}{Theoretical Astrophysics Group T-6, Mailstop B227, Los Alamos
National Laboratory, P.O. Box 1663, Los Alamos, NM 87545, USA}
\email{diehl@lanl.gov, statler@ohio.edu}

\begin{abstract}
We present an X-ray analysis of 54 normal elliptical galaxies in the {\it Chandra} archive and isolate their hot gas component from the contaminating point source emission, allowing us to conduct, for the first time, a morphological analysis on the gas alone. A comparison with optical images and photometry shows that the hot gas morphology has surprisingly little in common with the shape of the stellar distribution. We observe no correlation between optical and X-ray ellipticities in the inner regions where stellar mass dominates over dark matter. A shallow correlation would be expected if the gas had settled into hydrostatic equilibrium with the gravitational potential. Instead, observed X-ray ellipticities exceed optical ellipticities in many cases. We exclude rotation as the dominant factor to produce the gas ellipticities. The gas appears disturbed, and hydrostatic equilibrium is the exception rather than the rule. Nearly all hydrostatic models can be ruled out at 99\% confidence, based of their inability to reproduce the optical-X-ray correlation and large X-ray ellipticities. Hydrostatic models not excluded are those in which dark matter either dominates over stellar mass inside the inner half-light radius or has a prominently cigar-shaped distribution, both of which can be ruled out on other grounds. We conclude that, even for rather X-ray faint elliptical galaxies, the gas is at least so far out of equilibrium that it does not retain any information about the shape of the potential, and that X-ray derived radial mass profiles may be in error by factors of order unity.
\end{abstract}



\newcommand{\mycolspace}{}
\newcommand{\phone}{\phantom{1}}
\section{Introduction}\label{s3.introduction}

The launch of {\it Chandra} and {\it XMM-Newton} opened a new era in
our understanding of the hydrodynamic histories of galaxies and
clusters. Early spectral evidence from {\it ROSAT} and {\it ASCA}
\citep[e.g.][]{FabbianoROSATspectra, BuoteASCA} already suggested 
soft
diffuse gas and harder stellar point sources as the two dominant
components of the X-ray emission of normal elliptical galaxies. But
{\it Chandra} made it possible for the first time to spatially resolve
a significant fraction of the point source component into individual
sources, which are now believed to consist mainly of low-mass X-ray
binaries \citep[LMXBs, e.g.][and references
therein]{SarazinLMXBGC}. This has significantly contributed to the
understanding of the correlation between X-ray and blue
luminosity. The $L_{\rm X}$--$L_{\rm B}$ diagram shows a steep $L_{\rm
X}\propto L_{\rm B}^2$ relation at the gas-dominated group and cluster
scale, and gets shallower toward low X-ray luminosities for normal
elliptical galaxies due to the increasing importance of the point
source component \citep{OSullivanLxLb}. For these galaxies, the LMXB
emission severely contaminates the diffuse hot gas emission and
complicates efforts to reveal its spatial structure.

It has long been assumed that the hot interstellar medium (ISM) in
elliptical galaxies is in hydrostatic equilibrium with the underlying
gravitational potential \citep[e.g.][]{Forman1985}. The desire to make
this assumption is natural since it then gives us a powerful tool to
probe the host galaxy's mass distribution. As such, radial mass
profiles derived from observed X-ray pressure profiles are among the
strongest providers of evidence for the existence of massive dark
matter halos surrounding normal elliptical galaxies
\citep[e.g.][]{Forman1985,KilleenEinsteindarkmatter,FabbianoNGC507,
HumphreyDarkmatter,FukazawaMassprofiles} and have been found to 
be
consistent with those derived from gas kinematics in spiral galaxies
\citep[][and references therein]{SofueSpiralDarkmatter} and stellar
kinematics in ellipticals \citep[e.g.][]{StatlerDarkmatter}, implying
dark halos with mass profiles similar to those of isothermal
spheres. Unfortunately, stellar kinematics can only probe the dark
matter content out to a few optical radii. Farther out, one has to
rely on X-ray mass profiles, gravitational lensing
\citep{KeetonDarkmatter} or kinematical planetary nebula (PN) data
\citep{NapolitanoPN}. A recent study by \citet
{RomanowskyPNDarkmatter}
using PN data advocates a rather low dark matter contribution around some
elliptical galaxies, at odds with standard $\Lambda$-CDM simulations
\citep[e.g.][]{SpringelLargescalestruc}. This discrepancy may be
reconciled by appealing to radial orbits of the halo stars
\citep{DekelLostFound}, which underlines the difficulty of
interpreting PN kinematics.

The assumption of hydrostatic equilibrium in elliptical galaxies has
always been considered well founded and understood. That radial mass profiles
consistent with other methods can be derived from it indicates
that it is not seriously wrong. If hydrostatic equilibrium
holds precisely, the isophotes of the hot gas emission
should very nearly trace the projected potential isophotes, and may be
used to trace the total mass distribution. Buote and collaborators use this
approach to argue for a highly flattened triaxial dark matter halo in the
elliptical galaxy NGC~720 \citep{BuoteGeomtest,BuoteNGC720}. However,
no test has been devised to verify that hydrostatic equilibrium holds
precisely enough all the way through a galaxy to make this kind of inference
valid.

Indeed, substantial evidence is accumulating to the contrary. Combining
stellar kinematic information with X-ray observations of NGC~4472, both
\citet{CiottiXraymassproblems} and \citet{MathewsNGC4472orbits} independently
argue for a lack of hydrostatic equilibrium in this particular galaxy. 
\citet{NGC1700} argue that the extreme X-ray flattening of NGC~1700
provides a counterexample of an object that cannot be in hydrostatic
equilibrium, and suggest that it is rotationally flattened. Rotational
flattening might be expected for a variety of reasons
\citep{MathewsReview}. A large fraction of the hot gas in elliptical
galaxies is thought to come from stellar mass loss, and should carry
the stellar angular momentum. A fraction of the hot gas may also be
acquired externally during mergers, stripped off during close
encounters, or fall in from a circumgalactic gas reservoir
\citep{BrighentiCircumgas}. In each case, the gas should contain
significant angular momentum. In the standard cooling flow
scenario the gas should slowly flow inward, conserving angular
momentum and settling into a rotationally supported cooling disk
\citep{BrighentiDisks,BrighentiDisks2,KleyXraydisks}. However,
\citet{BregmanRotating} demonstrate, with {\it ROSAT} and {\it
Einstein} data for 6 elliptical galaxies, a lack of gas disk
signatures. They find ellipticities that generally do not exceed
values of $\sim 0.2$, whereas disk models predict ellipticities larger
than $\sim 0.5$.

Detailed {\it Chandra} and {\it XMM} studies of over two dozen
individual early-type galaxies are now published in the
literature. Many of these observations reveal systems that are
morphologically disturbed, with a large variety of suggested
causes. For example, \citet{Jon01} argue for the presence of shocks in
NGC~4636; \citet{Fin01} find evidence for interactions with the
central radio source in NGC~4374; and \citet{Machacek} suggest that
NGC~1404 is moving relative to the Fornax intracluster gas. Only a few
objects appear round and quiescent, such as NGC~4555
\citep{OSullivanNGC4555} or NGC~6482
\citep{PonmanNGC6482}. Nonetheless, the use 
of hydrostatic equilibrium as a fundamental assumption about the 
majority of early-type galaxies persists \citep[e.g.][]{FukazawaMassprofiles}. 

The main obstacle to understanding the true physical state of the hot
ISM in elliptical galaxies is the separation of the gas and unresolved
LMXB contributions to the diffuse emission, which is particularly
important for X-ray faint galaxies. So far, two main approaches have
been employed. (1) Resolved point sources are removed, and the
residual diffuse image is adaptively smoothed. It is then argued or
hoped that the contribution of unresolved point sources can be
neglected while interpreting the diffuse emission and that what is
shown is close to the gas morphology \citep[e.g.][]{BuoteNGC720}. (2)
The observation is separated into broad radial bins with sufficient
signal to extract and fit a spectrum with a two component model, thus
spectrally separating gas and LMXBs in the radial profile
\citep[e.g.][]{HumphreyDarkmatter}. Method (1) has the obvious
disadvantage that one does not know what one is looking at. One gains
no information on gas morphology in X-ray faint systems which are
known to be LMXB dominated. Method (2) allows some insight into the
radial distribution and extent of the gas but reveals nothing about
the true elliptical shape or asymmetric features.

This is the first paper in a series that analyzes data on 54
elliptical galaxies in the {\it Chandra} public archive. We
homogeneously reanalyze the observations and introduce a new technique
to isolate the gas from LMXBs. An adaptive binning technique
\citep{DiehlWVT,Cappellari} is then used to reveal the
morphology of the gas {\it alone}. We present a gas gallery and a
quantitative morphological analysis. For the first time, we are able to
test whether hydrostatic equilibrium generally holds in normal ellipticals, by
comparing gas and stellar isophotal ellipticities in the region where the
stellar mass is expected to dominate the potential. We show that the archive
sample fails the test. We also examine the evidence for rotational support of
the gas, and find it to be extremely weak. The subsequent papers in this
series will take up the origin of the observed gas morphology
\citep[][hereafter Paper II]{DiehlAGN} and address the importance of
central active galactic nuclei (AGN) in reheating the gas
\citep[][hereafter Paper III]{DiehlEntropy}.

The remainder of this paper is organized as follows. In
\S\ref{s3.data} we describe the details of the data reduction
pipeline, the new isolation technique and the method used to derive
ellipticity profiles. In \S\ref{s3.results} we present our comparison
between optical and X-ray properties and address the evidence for hydrostatic
equilibrium and 
rotational support. We discuss the implications of our findings in
\S\ref{s3.discussion}, before we summarize in \S\ref{s3.conclusions}.

\section{Data} \label{s3.data}
\subsection{Sample Selection and Pipeline Reduction}

We select all galaxies having non-grating ACIS-S
observations with approved exposure times longer than $10\ks$ in the
{\it Chandra} public archive of cycle 1-4 which are classified as E or E/S0 
in the Lyon--Meudon Extragalactic Database
\citep[LEDA;][]{LEDA}. We remove brightest cluster galaxies and
objects with purely AGN-dominated emission. This sample is almost
identical to that used in a previous paper \citep{Diehl_XGFP}, except
for the removal of two galaxies. We exclude the NGC~4782/NGC~4783
galaxy pair due to its ongoing merger and the dwarf elliptical NGC~1705
whose luminosity is two orders of magnitude smaller than that of the
next largest object. Our final sample consists of 54 early-type
galaxies, 34 of which are listed as members of a group in the
Lyon Group of Galaxies \citep[LGG,][]{LGG} catalog, with 19 identified
as the brightest group member. Eight are also identified as members of
X-ray-bright groups in the GEMS survey \citep{GEMS}, together with 5
additional galaxies that are not listed in the LGG catalog.

We apply a homogeneous data reduction pipeline to all observations
starting from their level~1event files using CIAO version 3.1 with calibration data
base 2.28. The basic data reduction steps follow the recommendations
according to {\it Chandra}'s ACIS data analysis
guide\footnote{http://cxc.harvard.edu/ciao/guides/acis\_data.html}. The
newest gain file is applied and adjusted for time-dependent gain
variations to account for the drift of the effective detector gains
with time caused by changes in the charge transfer
inefficiency. Observation-specific bad columns and pixels are removed
and each observation is restricted to its good time intervals. We
additionally filter each light curve by iteratively applying a
$2.5\sigma$ threshold to remove background flares. The remaining light
curve is then clipped at 20\% above the average count rate, to match
the standard of the Markevitch blank sky background
files\footnote{http://cxc.harvard.edu/cal/Acis/Cal\_prods/bkgrnd/acisbg/COOKBOOK}. Each
light curve is inspected and verified to be clean. If the entire
observation is affected by a very long flare that manifests itself as
an underlying ``ramp'' in the light curve, we flag the object and use
a local background spectrum in our spectral analysis instead of blank
sky fields. The background correction for the subsequent spatial
analysis is not taken from the blank sky background files, but rather
derived from surface brightness profile fits (see also
\S\ref{s3.isolation}) which will automatically correct for flux
offsets from residual flares. The remaining event list is filtered to
retain standard ASCA grades 0, 2, 3, 4, and 6 to optimize the
signal-to-background ratio. Cosmic ray afterglows, as flagged by the
CIAO tool {\it acis\_detect\_afterglow}, are removed for the
purpose of source detection to minimize the number of spurious
detections\footnote{http://cxc.harvard.edu/ciao/threads/acisdetectafterglow/}.
These photons are retained for the rest of the analysis. In practice,
we find that keeping or removing the afterglows makes no difference to
the results.

Our analysis is restricted to photon energies between $0.3$--$5\kev$,
maximizing the relative contribution of soft hot gas emission, while
avoiding the rise of the particle background at higher energies. All
quoted X-ray luminosities are restricted to this band. We split this
energy range further into a soft ($0.3$--$1.2\kev$) and a hard
($1.2$--$5.0\kev$) band. To create flux-calibrated images that take
the spatially dependent spectral changes into account, we create
mono-energetic exposure maps in steps of 7 in PI ($\sim 100\ev$). The
observation is split into $14.6\ev$-wide (the width of one PI channel)
individual images. A photon-flux-calibrated ``slice'' is created by
dividing this counts image by the energetically closest exposure
map. The sum of all individual slices represents the final photon flux
image. These flux-calibrated images allow an accurate flux
determination even in cases where a spectral analysis is impossible
due to the lack of sufficient signal.

\subsection{Isolating the X-ray Gas Emission}\label{s3.isolation}

To isolate the hot gas emission we follow the procedure outlined by
\citet{Diehl_XGFP}. A summary is repeated here with some additional
details, which are essential for the morphological analysis in this
paper.

We use the CIAO tool {\it wavdetect} to identify point sources, remove
regions enclosing 95\% of the source flux and refill the holes with
simulated Poisson counts to obtain an image of diffuse emission. We
determine uniform background values for the soft and hard bands by
extracting radial surface brightness profiles for the diffuse emission
from the calibrated photon-flux images and fitting them with $\beta$,
double-$\beta$ and S\'ersic models plus an additive constant
background. The sole purpose of these fits is to determine the
background value; all other fitting parameters are discarded for the
subsequent analysis. As the models generally yield slightly different
background values, we add half of this difference to the statistical 
background error. If there is insufficient signal to produce an
accurate fit, we compute the average surface brightness level outside
a 2.5 arcmin radius and use it as the background value. We then
subtract the spatially uniform backgrounds from the photon-flux
calibrated images to obtain the background corrected images of diffuse
emission for the soft and hard bands. Background errors are propagated 
into the final gas variance image
and taken into account in the subsequent analysis.

The hot gas emission is still contaminated by unresolved point sources
and the incompletely removed PSF wings of resolved point sources. These
contaminants must be modeled and removed
to isolate the hot gas emission alone. To do this, we split
the background corrected, photon-flux calibrated images into the soft
band $S$ and hard band $H$. Gas and point source components 
contribute
at different levels to each band, determined by their respective
softness ratios $\gamma$ and $\delta$. By expressing both bands as
linear combinations of the hot gas component $G$ and the unresolved
point source component $P$, we can solve this system of equations to
isolate the gas emission itself:
\begin{eqnarray}\label{e.linearcombination}
  S &=& \gamma P + \delta G, \\
  H &=& (1-\gamma) P + (1-\delta) G; \\ 
{\rm thus, }\quad  
G &=& {1-\gamma \over \delta-\gamma}\left[ S - \left({\gamma \over 1-
\gamma}
\right)\, H\right].
\end{eqnarray}
This decomposition depends on the softness ratios $\gamma$ and
$\delta$. To determine the softness of the unresolved point sources,
we first analyze the hardness ratios of resolved point sources. We
find no evidence for any spatial dependence or luminosity dependence
of their spectral properties. This is in agreement with studies by
\citet{Irwin03}, which suggest a universal nature of LMXBs, and allows
us to use the known spectral properties of resolved LMXBs as a
template for their unresolved counterparts. For each galaxy, we select
resolved point sources between $5\arcsec$ and 5 optical effective radii from the
center to avoid the influence of the central AGN, and to minimize
contributions from serendipitous background sources. We exclude high
luminosity sources ($>200$ counts) to ensure that the spectral fits
are driven by low-luminosity sources. If more than $10$ sources
fulfill these selection criteria, we fit an absorbed power law model
to the combined LMXB spectrum. We find that a narrow range of photon indices ($1.1$ to $1.8$) is adequate for the point source component in all of our galaxies). To get an equivalent point source
spectral model for galaxies without sufficient resolved point sources,
we simultaneously fit a power-law to all low-luminosity ($L_{\rm X}
\le 5\times 10^{37}\erg\,\s^{-1}$) LMXBs available in our complete
galaxy sample. This combined fit yields a photon index of $1.603$,
which we adopt as the representative spectral model for source-poor
galaxies. By integrating the adopted unabsorbed spectral model over
the soft and hard band, we derive the point source softness ratio
$\gamma$.

To find the softness ratio of the hot gas component, $\delta$, we
extract a spectrum from the diffuse emission within only the inner 3
optical effective radii of each galaxy, to avoid contamination by
intragroup or intracluster gas. We use an absorbed single temperature
APEC thermal plasma model for the hot gas and add the adopted
power-law model to represent unresolved point sources. We fix the
redshift parameter to the LEDA value, while temperature, metallicity
and normalizations are generally allowed to vary freely. For low
signal-to-noise spectra, we fix the metallicity to the solar
value. All described spectral fits are computed with the CIAO tool
{\it Sherpa} and corrected for Galactic absorption, with the hydrogen
column density fixed at the Galactic value for the line of sight, as
determined by the CIAO tool
{\it Colden}\footnote{http://cxc.harvard.edu/toolkit/colden.jsp}. This
decomposition technique removes both the unresolved point sources and
the PSF wings of resolved sources, due to their universal spectral nature.

To reliably interpret the sparse gas images and to reveal any spatial
features, we bin the gas images with an adaptive binning method using
weighted Voronoi tesselations
\citep{DiehlWVT,Cappellari}\footnote{http://www.phy.ohiou.edu/$\sim$diehl/WVT}
to achieve an approximately constant signal-to-noise ratio per bin of
$4$. In background dominated, very-low signal-to-noise regions, the
required bin size can get very large and the bins occasionally ``eat''
into the inner emission to accrete sufficient signal. To avoid this
effect, we restrict the maximum bin size for these objects, resulting
in a drop of signal-to-noise per bin in the outer regions.

A few things should be kept in mind regarding our new isolation
technique for the hot gas. First, we assume isothermal gas throughout
the galaxy, resulting in a spatially constant $\delta$ parameter. We
test the validity of this assumption by creating two-dimensional
temperature maps for systems with the highest signal-to-noise data and
correct the gas image for a spatially dependent $\delta$
value. Although most galaxies exhibit temperature gradients, the
corrections to the gas surface brightness are generally $<10\%$ and do not 
affect the overall gas morphology. For those objects that do have
sufficient signal to produce a spatially dependent temperature map,
the effects on ellipticity and asymmetry are negligible. Spatial
temperature gradients and inhomogeneities will be 
examined in more detail in Paper III. However, one should be aware
that strong localized temperature differences can potentially result
in an over-subtraction of hotter features and under-subtraction of
colder features. 

Second, the region influenced by the central PSF of an AGN may be
subtracted incorrectly, as the AGN softness ratio is most likely
different from that of the unresolved point source component. Thus, we
exclude the central regions from our analysis, and one should refrain
from interpreting the gas maps at the very center. A detailed analysis
of the central AGN and its effects on the overall morphology will be
presented separately in Paper II. One should also keep in mind that
possible direct X-ray jet signatures could be masked out in our
analysis, as jet knots would most likely be identified as point
sources and thus removed.\footnote{NGC 315 is the only object in our
sample with a conspicuous jet. We have verified by eye that the jet is
removed by our procedure.}

A drawback of our approach is the inevitable loss of signal. Since we
are subtracting a scaled version of the hard band from the soft band
to isolate the gas emission, we effectively subtract a part of the gas
flux. Even though our technique corrects for this missing flux, one
still loses a certain amount of signal, reducing the spatial
resolution in the adaptively binned gas image. The softness of the gas
comes close to that of the unresolved point sources for temperatures
exceeding $\sim 1.5\kev$. From that point on, both components become
increasingly indistinguishable and the signal-to-noise ratio of the
gas image can be very low, even if there is substantial gas
present. Fortunately, few objects in our sample have temperatures
exceeding $1\kev$.

The definite advantage of our algorithm is that we can properly
isolate the hot gas emission, without assumptions about the spatial
distribution of unresolved point sources, or their total flux. Even
the effect of a spatially variable point source detection limit (due
to variations in background levels and the changes in size and shape
of the point spread function) is compensated for by our algorithm.

To give the reader a better understanding of the relative
contributions of the hot gas and the resolved and unresolved point
source components, we list the relative fractions in Table
\ref{t.galaxyprop} for a radial range between $2.5\arcsec$ and 3
$J$-band effective radii ($R_{\rm J}$). This shows that for a large
subset of galaxies, a significant percentage of point sources is still
unresolved and contaminates the diffuse emission, disguising the true
gas morphology. In some cases (e.g. NGC~3115) the diffuse emission is
even consistent with being almost entirely due to unresolved point
sources.

\begin{deluxetable*}{lrrrrrrrr}
\tablewidth{0pt}
\tablecaption{{\it Chandra} X-ray properties\label{t.galaxyprop}}
\tablehead{
\colhead{Name} 
& \colhead{ObsId} 
& \colhead{$\tau$\tablenotemark{a}} 
&  \colhead{$f_{\rm Gas}$\tablenotemark{b}} 
& \colhead{$f_{\rm unres}$\tablenotemark{b}} 
& \colhead{$f_{\rm resol}$\tablenotemark{b}} 
& \colhead{$L_{\rm X,Gas}$\tablenotemark{c}} 
&\colhead{$\epsilon_{\rm X}$\tablenotemark{d}} 
& \colhead{${\rm PA}_{\rm X}$\tablenotemark{d}} }
\startdata
IC1262 &  $ 2018 $  &  $           31 $  &  $  58 \pm  15\% $  &  $  41 \pm  15\% $  &  $   1 \pm  \phantom{1}   1\% $  &  $  2.0\pm 1.7\times 10^{43} $  & $0.43 \pm 0.09 $ &  $69 \pm 11$\\
IC1459 &  $ 2196 $  &  $           52 $  &  $  58 \pm  10\% $  &  $  12 \pm  10\% $  &  $  31 \pm  \phantom{1}   1\% $  &  $  4.3\pm 3.2\times 10^{40} $  & $  0.16 \pm 0.08 $ & $ 32  \pm 20 $ \\
IC4296 &  $ 3394 $  &  $            24 $  &  $  74 \pm  10\% $  &  $   7 \pm  \phantom{1}   9\% $  &  $  18 \pm  \phantom{1}   1\% $  &  $  1.1\pm 0.4\times 10^{41} $  & $ 0.28 \pm 0.06 $ & $ 55 \pm 14  $ \\
NGC0193 &  $ 4053 $  &  $           29 $  &  $  85 \pm  13\% $  &  $  10 \pm  \phantom{1}   9\% $  &  $   5 \pm  \phantom{1}   1\% $  &  $  2.5\pm 0.8\times 10^{41} $  & \nodata & \nodata \\
NGC0315 &  $ 4156 $  &  $           52 $  &  $  58 \pm  10\% $  &  $  32 \pm  10\% $  &  $  10 \pm  \phantom{1}   1\% $  &  $  9.4\pm 3.4\times 10^{40} $  & $ 0.13 \pm 0.16 $ & $ 14 \pm 32 $ \\
NGC0383 &  $ 2147 $  &  $           43 $  &  $  49 \pm  10\% $  &  $  35 \pm  10\% $  &  $  16 \pm  \phantom{1}   1\% $  &  $ <  7.5\times 10^{41} $  & $ 0.23 \pm 0.09 $ & $114 \pm 25 $ \\
NGC0404 &  $ 870 $  &  $           24 $  &  $  94 \pm  43\% $  &  $   0 \pm  42\% $  &  $   6 \pm  \phantom{1}   2\% $  &  $ <  2.1\times 10^{38} $  & \nodata & \nodata \\
NGC0507 &  $ 317 $  &  $           26 $  &  $  69 \pm  13\% $  &  $  30 \pm  13\% $  &  $   1 \pm  \phantom{1}   1\% $  &  $ >  5.7\times 10^{42} $  & $0.13 \pm 0.17$ & $91 \pm 33 $ \\
NGC0533 &  $ 2880 $  &  $           35 $  &  $  89 \pm  19\% $  &  $   9 \pm  18\% $  &  $   2 \pm  \phantom{1}   1\% $  &  $  9.6\pm 3.5\times 10^{41} $  & $0.37 \pm 0.05 $ & $ 26 \pm 3 $ \\
NGC0720 &  $ 492 $  &  $           34 $  &  $  74 \pm  10\% $  &  $  15 \pm  \phantom{1}   9\% $  &  $  12 \pm  \phantom{1}   1\% $  &  $  9.3\pm 2.7\times 10^{40} $  & $ 0.09 \pm 0.07 $ & $ 75  \pm 28  $ \\
NGC0741 &  $ 2223 $  &  $           29 $  &  $  58 \pm  11\% $  &  $  21 \pm  11\% $  &  $  22 \pm  \phantom{1}   1\% $  &  $  3.2\pm 1.3\times 10^{41} $  & $ 0.24 \pm 0.05 $ & $ 162 \pm 13  $ \\
NGC0821 &  $ 4006 $  &  $           13 $  &  $  17 \pm  23\% $  &  $   5 \pm  87\% $  &  $  78 \pm  54\% $  &  $ <  3.3\times 10^{40} $  & \nodata & \nodata \\
NGC1132 &  $ 801 $  &  $           12 $  &  $  83 \pm  16\% $  &  $  16 \pm  16\% $  &  $   1 \pm  \phantom{1}   1\% $  &  $ >  9.1\times 10^{42} $  & $ 0.08 \pm 0.24$  & $40 \pm 46 $  \\
NGC1265 &  $ 3237 $  &  $           82 $  &  $  15 \pm  \phantom{1}   8\% $  &  $  38 \pm  \phantom{1}   8\% $  &  $  47 \pm  \phantom{1}   5\% $  &  $ <  1.1\times 10^{42} $  & \nodata & \nodata \\
NGC1316 &  $ 2022 $  &  $           26 $  &  $  78 \pm  13\% $  &  $   5 \pm  13\% $  &  $  18 \pm  \phantom{1}   1\% $  &  $  5.7\pm 2.1\times 10^{40} $  & $ 0.33 \pm 0.06 $  & $ 14 \pm 13 $  \\
NGC1399 &  $ 319 $  &  $           24 $  &  $  85 \pm  23\% $  &  $   7 \pm  23\% $  &  $   8 \pm  \phantom{1}   1\% $  &  $ >  7.9\times 10^{41} $  & $ 0.34 \pm 0.04 $ & $ 179   \pm 7 $ \\
NGC1404 &  $ 2942 $  &  $           29 $  &  $  94 \pm  16\% $  &  $   1 \pm  16\% $  &  $   5 \pm  \phantom{1}   1\% $  &  $  1.7\pm 0.4\times 10^{41} $  & $ 0.06 \pm 0.07 $ & $ 63  \pm 26  $ \\
NGC1407 &  $ 791 $  &  $           43 $  &  $  70 \pm  10\% $  &  $  18 \pm  10\% $  &  $  12 \pm  \phantom{1}   1\% $  &  $  1.0\pm 0.3\times 10^{41} $  & $ 0.13 \pm 0.06 $ & $  165 \pm 16  $ \\
NGC1549 &  $ 2077 $  &  $           20 $  &  $  52 \pm  10\% $  &  $  18 \pm  \phantom{1}   9\% $  &  $  31 \pm  \phantom{1}   2\% $  &  $ >  2.0\times 10^{40} $  & \nodata & \nodata \\
NGC1553 &  $ 783 $  &  $           19 $  &  $  61 \pm  13\% $  &  $  24 \pm  13\% $  &  $  15 \pm  \phantom{1}   1\% $  &  $  2.8\pm 2.6\times 10^{40} $  & $0.57 \pm 0.14 $  & $ 90 \pm 15$ \\
NGC1600 &  $ 4283/4371 $  &  $           49 $  &  $  79 \pm  13\% $  &  $  17 \pm  12\% $  &  $   4 \pm  \phantom{1}   1\% $  &  $ >  1.2\times 10^{42} $  & $ 0.24 \pm 0.09 $ & $ 178 \pm 17  $ \\
NGC1700 &  $ 2069 $  &  $           39 $  &  $  85 \pm  14\% $  &  $   7 \pm  14\% $  &  $   9 \pm  \phantom{1}   1\% $  &  $ >  3.2\times 10^{41} $  & $ 0.20 \pm 0.06 $ & $ 89  \pm 12  $ \\
NGC2434 &  $ 2923 $  &  $           25 $  &  $  69 \pm  \phantom{1}   9\% $  &  $   9 \pm  \phantom{1}   6\% $  &  $  21 \pm  \phantom{1}   1\% $  &  $  2.6\pm 2.0\times 10^{40} $  & \nodata & \nodata \\
NGC2865 &  $ 2020 $  &  $           25 $  &  $  34 \pm  21\% $  &  $  60 \pm  19\% $  &  $   6 \pm  \phantom{1}   2\% $  &  $ <  9.9\times 10^{40} $  & \nodata & \nodata \\
NGC3115 &  $ 2040 $  &  $           14 $  &  $   2 \pm  \phantom{1}   9\% $  &  $  35 \pm  \phantom{1}   8\% $  &  $  63 \pm  \phantom{1}   5\% $  &  $ <  8.7\times 10^{39} $  & \nodata & \nodata \\
NGC3377 &  $ 2934 $  &  $           39 $  &  $  15 \pm  22\% $  &  $   0 \pm  15\% $  &  $  85 \pm  10\% $  &  $ <  6.1\times 10^{39} $  & \nodata & \nodata \\
NGC3379 &  $ 1587 $  &  $           30 $  &  $  14 \pm  \phantom{1}   4\% $  &  $  11 \pm  \phantom{1}   4\% $  &  $  75 \pm  \phantom{1}   2\% $  &  $ <  6.3\times 10^{39} $  & \nodata & \nodata \\
NGC3585 &  $ 2078 $  &  $           35 $  &  $  32 \pm  \phantom{1}   7\% $  &  $  19 \pm  \phantom{1}   5\% $  &  $  48 \pm  \phantom{1}   3\% $  &  $ >  4.2\times 10^{39} $  & \nodata & \nodata \\
NGC3923 &  $ 1563 $  &  $           16 $  &  $  74 \pm  \phantom{1}   9\% $  &  $  11 \pm  \phantom{1}   8\% $  &  $  15 \pm  \phantom{1}   1\% $  &  $  4.3\pm 1.3\times 10^{40} $  & $ 0.15 \pm 0.10 $ & $ 44  \pm 25  $ \\
NGC4125 &  $ 2071 $  &  $           63 $  &  $  80 \pm  13\% $  &  $   1 \pm  13\% $  &  $  19 \pm  \phantom{1}   1\% $  &  $  7.2\pm 2.7\times 10^{40} $  & $ 0.24 \pm 0.11 $ & $  71 \pm 10  $ \\
NGC4261 &  $ 834 $  &  $           31 $  &  $  63 \pm  11\% $  &  $  18 \pm  10\% $  &  $  19 \pm  \phantom{1}   1\% $  &  $  4.8\pm 1.1\times 10^{40} $  & $ 0.08 \pm 0.08 $ & $ 108 \pm 35  $ \\
NGC4365 &  $ 2015 $  &  $           40 $  &  $  44 \pm  \phantom{1}   8\% $  &  $  14 \pm  \phantom{1}   8\% $  &  $  41 \pm  \phantom{1}   1\% $  &  $ >  3.8\times 10^{40} $  & \nodata & \nodata \\
NGC4374 &  $ 803 $  &  $           28 $  &  $  86 \pm  18\% $  &  $   5 \pm  18\% $  &  $   9 \pm  \phantom{1}   1\% $  &  $  5.9\pm 1.3\times 10^{40} $  & $ 0.38 \pm 0.06 $ & $ 116 \pm 7   $ \\
NGC4406 &  $ 318 $  &  $           14 $  &  $  89 \pm  12\% $  &  $  10 \pm  12\% $  &  $   2 \pm  \phantom{1}   1\% $  &  $ >  1.0\times 10^{42} $  & $ 0.17 \pm 0.10 $ & $ 113 \pm 24  $ \\
NGC4472 &  $ 321 $  &  $           35 $  &  $  83 \pm  16\% $  &  $  12 \pm  16\% $  &  $   5 \pm  \phantom{1}   1\% $  &  $ >  8.5\times 10^{41} $  & $ 0.12 \pm 0.04 $ & $ 136  \pm 13  $ \\
NGC4494 &  $ 2079 $  &  $           19 $  &  $   0 \pm  \phantom{1}   2\% $  &  $  43 \pm  10\% $  &  $  57 \pm  \phantom{1}   5\% $  &  $ <  2.1\times 10^{40} $  & \nodata & \nodata \\
NGC4526 &  $ 3925 $  &  $           38 $  &  $  55 \pm  12\% $  &  $   8 \pm  10\% $  &  $  37 \pm  \phantom{1}   2\% $  &  $  8.8\pm 7.5\times 10^{39} $  & $0.07 \pm 0.15 $  & $1 \pm 43$  \\
NGC4552 &  $ 2072 $  &  $           53 $  &  $  80 \pm  10\% $  &  $   4 \pm  10\% $  &  $  16 \pm  \phantom{1}   1\% $  &  $  2.1\pm 1.2\times 10^{40} $  & $ 0.21 \pm 0.03 $ & $  7  \pm 4   $ \\
NGC4555 &  $ 2884 $  &  $           27 $  &  $  73 \pm  20\% $  &  $  19 \pm  18\% $  &  $   9 \pm  \phantom{1}   2\% $  &  $ >  2.3\times 10^{41} $  & \nodata & \nodata \\
NGC4564 &  $ 4008 $  &  $           17 $  &  $  72 \pm  23\% $  &  $  17 \pm  15\% $  &  $  11 \pm  \phantom{1}   2\% $  &  $ >  2.0\times 10^{39} $  & \nodata & \nodata \\
NGC4621 &  $ 2068 $  &  $           24 $  &  $  42 \pm  11\% $  &  $   7 \pm  \phantom{1}   9\% $  &  $  50 \pm  \phantom{1}   4\% $  &  $  1.1\pm 0.9\times 10^{40} $  & \nodata & \nodata \\
NGC4636 &  $ 323 $  &  $           46 $  &  $  95 \pm  17\% $  &  $   3 \pm  17\% $  &  $   3 \pm  \phantom{1}   1\% $  &  $  2.7\pm 2.0\times 10^{41} $  & $ 0.35 \pm 0.04 $ & $  9  \pm  3  $ \\
NGC4649 &  $ 785 $  &  $           31 $  &  $  80 \pm  11\% $  &  $  11 \pm  11\% $  &  $   9 \pm  \phantom{1}   1\% $  &  $  1.3\pm 0.3\times 10^{41} $  & $ 0.08 \pm 0.03 $ & $  95 \pm  27 $ \\
NGC4697 &  $ 784/4727/4728 $  &  $          112 $  &  $  57 \pm  \phantom{1}   6\% $  &  $   6 \pm  \phantom{1}   5\% $  &  $  38 \pm  \phantom{1}   1\% $  &  $ >  3.5\times 10^{40} $  & $ 0.20 \pm 0.16 $ & $ 80  \pm  23 $ \\
NGC5018 &  $ 2070 $  &  $           25 $  &  $  61 \pm  11\% $  &  $  21 \pm  \phantom{1}   9\% $  &  $  18 \pm  \phantom{1}   2\% $  &  $ <  1.9\times 10^{41} $  & \nodata & \nodata \\
NGC5044 &  $ 798 $  &  $           20 $  &  $ 100 \pm  18\% $  &  $   0 \pm  18\% $  &  $   0 \pm  \phantom{1}   1\% $  &  $  2.6\pm 0.8\times 10^{42} $  & $ 0.41 \pm 0.08 $ & $ 27  \pm 10   $ \\
NGC5102 &  $ 2949 $  &  $           34 $  &  $  50 \pm  22\% $  &  $   4 \pm  13\% $  &  $  46 \pm  \phantom{1}   7\% $  &  $ <  1.6\times 10^{39} $  & \nodata & \nodata \\
NGC5171 &  $ 3216 $  &  $           34 $  &  $  83 \pm  25\% $  &  $  13 \pm  20\% $  &  $   4 \pm  \phantom{1}   1\% $  &  $ >  2.7\times 10^{42} $  & \nodata & \nodata \\
NGC5532 &  $ 3968 $  &  $           48 $  &  $  28 \pm  \phantom{1}   6\% $  &  $  24 \pm  \phantom{1}   5\% $  &  $  48 \pm  \phantom{1}   2\% $  &  $ <  8.7\times 10^{41} $  & \nodata & \nodata \\
NGC5845 &  $ 4009 $  &  $           30 $  &  $  57 \pm  30\% $  &  $  41 \pm  19\% $  &  $   2 \pm  \phantom{1}   3\% $  &  $ <  5.2\times 10^{40} $  & \nodata & \nodata \\
NGC5846 &  $ 788 $  &  $           24 $  &  $  95 \pm  16\% $  &  $   4 \pm  16\% $  &  $   2 \pm  \phantom{1}   1\% $  &  $  3.9\pm 0.9\times 10^{41} $  & $ 0.22 \pm 0.05 $ & $  175 \pm 8   $ \\
NGC6482 &  $ 3218 $  &  $           19 $  &  $  90 \pm  13\% $  &  $   9 \pm  13\% $  &  $   1 \pm  \phantom{1}   1\% $  &  $  1.7\pm 1.3\times 10^{42} $  & $ 0.18 \pm 0.09 $ & $ 36  \pm 23  $ \\
NGC7052 &  $ 2931 $  &  $            9 $  &  $  90 \pm  14\% $  &  $   6 \pm  10\% $  &  $   4 \pm  \phantom{1}   1\% $  &  $ >  1.1\times 10^{41} $  & $ 0.26 \pm 0.12 $ & $ 54  \pm 23  $ \\
NGC7618 &  $ 802 $  &  $           11 $  &  $  85 \pm  11\% $  &  $  14 \pm  11\% $  &  $   1 \pm  \phantom{1}   1\% $  &  $  2.3\pm 0.9\times 10^{42} $  & $0.38 \pm 0.14$  & $99 \pm 16 $  \\
\enddata
\tablenotetext{a}{Effective {\it Chandra} exposure time in $\ks$ after
removal of flares.}
\tablenotetext{b}{Percentages of observed photon fluxes for 
the hot gas, unresolved and resolved point sources, integrated between
 $2.5\arcsec$ and $3R_{\rm J}$ .}
\tablenotetext{c}{Total X-ray gas luminosity in ${\rm ergs\,s^{-1}}$ 
for the $0.3-5\kev$ band.}
\tablenotetext{d}{X-ray gas ellipticity and position angle, evaluated 
between $0.6$ and $0.9\,R_{\rm J}$.}
\end{deluxetable*}

\subsection{X-ray Gas Luminosity}

To calculate the total X-ray gas luminosity, we bin the
calibrated gas images into adaptively sized circular annuli and produce radial surface
brightness profiles, as described in \citet{Diehl_XGFP}. X-ray surface
brightness profiles of galaxy clusters and groups are generally well
described by $\beta$ or double-$\beta$ profiles with $\beta$ values
between $\sim 0.5-1.0$. Our $\beta$ model fits indicate that the gas
profiles of normal ellipticals are generally shallower. For 28 out of
45 galaxies with sufficient signal to constrain the fit, the best fit
$\beta$ values are $<0.5$, consistent with the known relation between
gas temperature and $\beta$ for groups and clusters
\citep[e.g.][]{VoitModentropy}.

However, a $\beta$ value below $0.5$ yields infinite total flux when
extrapolated to large radii. As this renders the $\beta$ models
unusable for determining total luminosities, we adopt the well-known
S\'ersic models instead. The S\'ersic model has the advantage of
yielding finite fluxes and produces equivalently good fits. The values
for our best S\'ersic model fits are published in an earlier paper
\citep{Diehl_XGFP}. We derive X-ray gas luminosities by summing the
calibrated gas images over the field of view, and use the model fits
to correct for missing flux outside the field of view. The
luminosities listed in Table \ref{t.galaxyprop} are also corrected for
absorption effects by multiplying a correction factor, derived from
the best spectral fit, which we integrate over energy with and without absorption
by the Galactic neutral hydrogen. The error bars for luminosities can
get quite large for objects with very wide gas emission, where a
significant fraction of flux is derived from the uncertain
extrapolation to large radii. The quoted uncertainties in Table
\ref{t.galaxyprop} also include systematic errors associated with
uncertainties in the $\delta$ and $\gamma$ parameters from section
\ref{s3.isolation}, as well as the uncertainty in our adopted
distances \citep[for more details, see][]{Diehl_XGFP}. In cases where
the uncertainties are larger than the actual values due to
uncertainties in the extrapolation or where we have insufficient
signal for a surface brightness model fit, we determine the luminosity
by summing the flux in the field of view. If the summed flux
represents a detection at a $>3\sigma$ level, we report the $3\sigma$
lower bound as a low limit in Table \ref{t.galaxyprop}. If this is not
the case, we report the $3\sigma$ upper bound as an upper limit.

\subsection{X-ray Ellipticity and Position Angle Profiles}

The extremely sparse nature of X-ray data generally prohibits the use
of isophote fitting techniques commonly employed with optical
data. Moreover, because our gas-only images are obtained from scaled
differences of the hard and soft bands, they contain many individual
pixels with large negative counts, which can render standard
algorithms numerically unstable. To avoid this problem, we fit
isophotes to the adaptively binned gas images, adapting techniques
used in $N$-body simulations.

We populate each bin randomly with a number of pseudo counts
(particles) such that the expected $\sqrt{N}$ Poisson fluctuations
match the signal-to-noise ratio in the bin. The pseudo count fluxes
(masses) are chosen to give the correct flux in each bin. Isophotes
are then fitted to the pseudo count distribution using an iterative
algorithm that diagonalizes the second-moment tensor in a thin
elliptical ring and manipulates the axis ratio until the eigenvalues
match those for a constant density ring of the same shape
\citep{NGC1700}. Twenty random realizations are run, and the
distribution of complex ellipticities is used to find the mean
isophotal ellipticity, major axis position angle, and errors at each
radius. 

\begin{figure}
\plotone{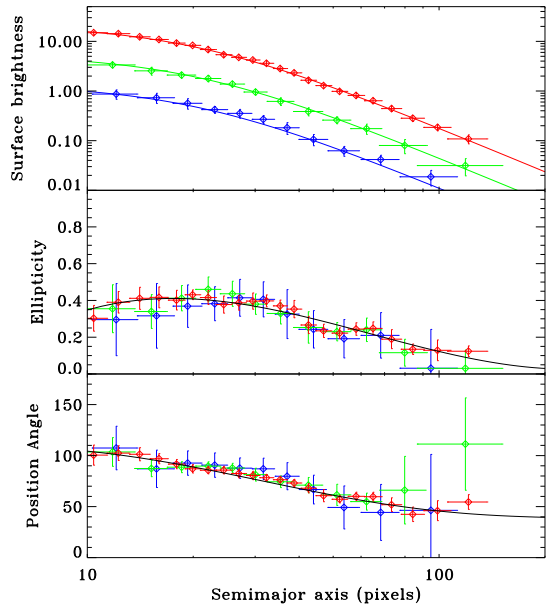}
\caption{Tests of isophote fitting for simulated data with $40,000$ (red),
$10,000$ (green), and $2,500$ (blue) counts from gas. Gas emission is
separated from point source emission by identifying and removing resolved sources and differencing
soft- and hard-band images to remove unresolved sources (\S\ 2.2) and PSF wings. Isophotes are fitted to
the adaptively binned gas image as described in \S\ 2.4. Top panel shows
gas surface brightness (in counts/pixel). Middle and bottom panels show
isophotal ellipticity and major axis poisition angle. Smooth curves indicate
the input profiles.
\label{f.isophote_test}}
\end{figure}

Tests show that this technique is able to robustly recover simulated isophotal
profiles. An example is shown in figure \ref{f.isophote_test}. Here we have
simulated soft- and hard-band images from separate gas and point
source contributions. Both gas and point sources follow different radial 
surface brightness models and have very different isophotal shapes. 
The point sources are distributed as a deVaucouleurs
profile with an effective radius of $50$ pixels with a constant ellipticity of $0.1$ 
and major-axis position angle of $-10^\circ$. We draw the point source luminosities 
from a typical power-law luminosity function with a logarithmic slope of $-2.0$. We then 
compute the {\it Chandra} point spread function with the CIAO tool ``mkpsf'' 
at the source location and finally simulate Poisson counts. 
The gas on the other hand follows a $\beta$ profile with a core radius of $20$ 
pixels and $\beta=2/3$. The more complicated gas ellipticity and position angle 
profiles are shown by the smooth curves in the middle and bottom panels of 
Figure \ref{f.isophote_test}. We assume typical softness ratios of
$\gamma = 0.5$ and $\delta = 0.9$ and choose the luminosities such that 60\% of the counts in the
simulated full-band image are from gas. This percentage is representative of
the lowest end of the range for the objects in Table 1 for which isophote
fitting is possible. A flat background of $0.01$ counts/pixel is added to the
soft and hard images before simulating discrete counts. We then go through our
entire analysis pipeline: detecting point sources, removing resolved point sources, 
differencing the images to extract the gas map, adaptively bin, and fit isophotes as described
above. Figure \ref{f.isophote_test} shows the results for $40,000$, $10,000$, and $2,500$
gas counts for a comparable radial range as is used for the real data. The isophotal profiles are 
recovered to within the statistical errors. Clearly, at least a few thousand counts in gas are necessary to
recover accurate ellipticities. Only in cases where the entire central region 
is identified as consisting of several point sources and is completely blocked out, 
we note a systematic deviation from the input profiles. In our data this region is found to be within the central $10$ pixels. Thus, we do not interpret the central 
$5\arcsec$ ($10$ pixels) in our isophotal profiles. Furthermore, accurate isophotes can be fit
only where the gas surface brightness is $\gtrsim 2$ times the background. 
As a result, the isophotes are completely insensitive to errors in the background
level.

\subsection{Optical Data}

Optical data for the sample galaxies are given in Table
\ref{t.optprop}. We adopt the effective $J$-band ellipticity
$\epsilon_{\rm J}$, position angle ${\rm PA}_{\rm J}$ and half-light
radius $R_{\rm J}$ (Table \ref{t.optprop}), as well as the absolute
$K$ magnitude from the 2MASS extended source catalog
\citep{2MASS}. These ellipticities and position angles are extracted
at a constant magnitude level, approximately $3\sigma$ above
background noise level, generally translating to an extraction radius
between 20--40 arcsec.

In addition, we extract $R$-band images from the second epoch of the
Digitized Sky Survey (DSS-2R) and make use of publicly available
optical surface photometry from the literature. In particular, we
adopt $B$ (20 galaxies), $V$ (20), and $I$ (19) photometry from
\citet{GoudfrooijPhot}\footnote{PA profiles from
\citet{GoudfrooijPhot} have been corrected for mirror-image flips by
matching the profiles to other published photometry and DSS-2R
images.}, $V$ (21), $R$ (25), and $I$ (23) from \citet{BenderPhot},
$U$ (13), $B$ (15), and $R$ (14) from \citet{PeletierPhot} and $F814W$
(9) photometry from \citet{MDMPhot}. Refer to Table \ref{t.optprop}
for details. We use the optical photometry to extract effective
optical position angles ${\rm PA}_{\rm opt}$ and ellipticities
$\epsilon_{\rm opt}$ between $0.6$--$0.9\,R_{\rm J}$.

We also query the Hyperleda \citep{Hyperleda} data base to derive
average rotational velocities between $0.6$--$0.9\,R_{\rm J}$. In
cases where Hyperleda does not provide the data in electronic form, we
read off approximate values from published kinematic profile
plots. The list of references that are used to derive the adopted
rotational velocity is given in Table \ref{t.optprop}. This rotational
velocity is closely related to the maximal rotational velocity but is
generally a better defined quantity, as it is independent of the
observational cutoff of the available kinematic data.

\tabletypesize{\scriptsize}

\begin{deluxetable*}{lrcrrrrcrcrccrrc}
\tablewidth{0pt}
\tablecaption{Optical properties\label{t.optprop}}
\tablehead{
\colhead{} & \colhead{} & \colhead{} & 
\multicolumn{4}{c}{2MASS\tablenotemark{b}} & \colhead{} & 
\colhead{LEDA\tablenotemark{c}} & \colhead{} &
\multicolumn{2}{c}{Hyperleda\tablenotemark{d}} & \colhead{} & 
\multicolumn{3}{c}{Photometry\tablenotemark{e}} \\
\cline{4-7} \cline{9-9} \cline{11-12} \cline{14-16}\\
\colhead{Name} & \colhead{$D$\tablenotemark{a}} & \colhead{} &
\colhead{$M_{\rm K}$} & \colhead{$R_{\rm J}$} & 
\colhead{$\epsilon_{\rm J}$} & \colhead{${\rm PA}_{\rm J}$} &  
\colhead{\mycolspace} &
\colhead{$L_{\rm B}$} &  \colhead{\mycolspace} &
\colhead{$v_{\rm rot}$} & \colhead{Ref.} &  \colhead{\mycolspace} &
\colhead{$\epsilon_{\rm opt}$} & \colhead{${\rm PA}_{\rm opt}$} & 
\colhead{Ref.\tablenotemark{e}}}
\startdata
IC1262 &  $ 143.8 \pm  21.6 $  &  &  $ -25.43 \pm   0.33 $  &  $ 14.4 $  &  $ 0.34 $  &  $  85 $  &  &  $  5.4\pm 1.7\times 10^{10} $  &  &    \nodata    &  \nodata  &  &  \nodata  &  \nodata  &  \nodata   \\
IC1459 &  $  29.2 \pm \phantom{1}  3.8 $  &  &  $ -25.53 \pm   0.28 $  &  $ 29.1 $  &  $ 0.19 $  &  $  42 $  &  &  $  4.7\pm 2.3\times 10^{10} $  &  &  \nodata  &  \nodata  &  &  $ 0.27  \pm 0.01 $  &  $  36.6  \pm 0.7 $  & T    \\
IC4296 &  $  51.6 \pm \phantom{1}   7.7 $  &  &  $ -26.06 \pm   0.33 $  &  $ 25.5 $  &  $ 0.06 $  &  $  70 $  &  &  $  1.3\pm 0.4\times 10^{11} $  &  &  \nodata  &  \nodata  &  &  $ 0.13  \pm 0.01 $  &  $  65.9  \pm 1.0 $  &  U   \\
NGC0193 &  $  60.8 \pm \phantom{1}   9.1 $  &  &  $ -24.71 \pm   0.33 $  &  $ 14.5 $  &  $ 0.20 $  &  $  55 $  &  &  $  3.0\pm 1.0\times 10^{10} $  &  &  \nodata  &  \nodata  &  &  \nodata  &  \nodata  &  \nodata   \\
NGC0315 &  $  71.9 \pm  10.8 $  &  &  $ -26.33 \pm   0.33 $  &  $ 22.9 $  &  $ 0.22 $  &  $  50 $  &  &  $  1.4\pm 0.4\times 10^{11} $  &  &  $ 111 $  & A &  &  $ 0.27  \pm 0.01  $  &  $  42.5  \pm 0.5 $  & V    \\
NGC0383 &  $  73.2 \pm  11.0 $  &  &  $ -25.84 \pm   0.33 $  &  $ 17.8 $  &  $ 0.16 $  &  $  15 $  &  &  $ 1.0\pm 0.5\times 10^{11} $  &  &  $  75 $  & A &  &  \nodata  &  \nodata  &  \nodata   \\
NGC0404 &  $   3.3 \pm  \phantom{1}  0.2 $  &  &  \nodata  &  \nodata  &  \nodata  &  \nodata  &  &  $  7.0\pm 1.0\times 10^{8\phantom{0}} $  &  &  \nodata  &  \nodata  &  &  \nodata  &  \nodata  &  \nodata   \\
NGC0507 &  $  71.7 \pm  10.8 $  &  &  $ -25.98 \pm   0.33 $  &  $ 26.1 $  &  $ 0.12 $  &  $  75 $  &  &  $  1.1\pm 0.5\times 10^{11} $  &  &  \nodata  &  \nodata  &  &  \nodata  &  \nodata  &  \nodata   \\
NGC0533 &  $  77.6 \pm  11.6 $  &  &  $ -26.01 \pm   0.33 $  &  $ 25.2 $  &  $ 0.20 $  &  $  60 $  &  &  $  1.0\pm 0.5\times 10^{11} $  &  &  $  20 $  & B &  &  \nodata  &  \nodata  &  \nodata   \\
NGC0720 &  $  27.7 \pm  \phantom{1}  2.2 $  &  &  $ -24.94 \pm   0.17 $  &  $ 27.4 $  &  $ 0.41 $  &  $ 145 $  &  &  $  3.6\pm 1.4\times 10^{10} $  &  &  \nodata  &  \nodata  &  &  $ 0.41  \pm 0.01 $  &  $ 140.7  \pm 0.1 $  & T    \\
NGC0741 &  $  79.0 \pm  11.8 $  &  &  $ -26.19 \pm   0.33 $  &  $ 25.9 $  &  $ 0.38 $  &  $  95 $  &  &  $  1.6\pm 0.5\times 10^{11} $  &  &  $ 133 $  & C &  &  $ 0.18 \pm 0.01   $  &  $  87.8  \pm 1.2 $  & V    \\
NGC0821 &  $  24.1 \pm  \phantom{1}  1.9 $  &  &  $ -24.01 \pm   0.17 $  &  $ 23.9 $  &  $ 0.24 $  &  $  25 $  &  &  $  2.5\pm 0.9\times 10^{10} $  &  &  $  99 $  & D &  &  $ 0.38 \pm 0.01  $  &  $  32.3 \pm 0.6 $  &  W  \\
NGC1132 &  $  98.2 \pm  14.7 $  &  &  $ -25.70 \pm   0.33 $  &  $ 19.8 $  &  $ 0.30 $  &  $ 140 $  &  &  $  8.4\pm 4.2\times 10^{10} $  &  &  \nodata  &  \nodata  &  &  \nodata  &  \nodata  &  \nodata   \\
NGC1265 &  $ 109.5 \pm  16.4 $  &  &  \nodata  &  \nodata  &  \nodata  &  \nodata  &  &  $  2.1\pm 0.9\times 10^{11} $  &  &  \nodata  &  \nodata  &  &  \nodata  &  \nodata  &  \nodata   \\
NGC1316 &  $  21.5 \pm  \phantom{1}  1.7 $  &  &  $ -26.07 \pm   0.17 $  &  $ 49.8 $  &  $ 0.30 $  &  $  52 $  &  &  $  9.8\pm 9.2\times 10^{10} $  &  &  $ 134 $  & E &  &  \nodata  &  \nodata  &  \nodata   \\
NGC1399 &  $  20.0 \pm  \phantom{1}  1.5 $  &  &  $ -25.19 \pm   0.16 $  &  $ 36.9 $  &  $ 0.04 $  &  $ 150 $  &  &  $  4.9\pm 1.6\times 10^{10} $  &  &  $  31 $  & F &  &  $ 0.10 \pm 0.01   $  &  $ 107.4  \pm 1.0 $  & T    \\
NGC1404 &  $  21.0 \pm  \phantom{1}  1.8 $  &  &  $ -24.79 \pm   0.19 $  &  $ 19.3 $  &  $ 0.12 $  &  $ 163 $  &  &  $  3.2\pm 0.7\times 10^{10} $  &  &  $  89 $  & F &  &  $ 0.14 \pm 0.01   $  &  $ 160.1  \pm 1.2 $  &  T   \\
NGC1407 &  $  28.8 \pm  \phantom{1}  3.5 $  &  &  $ -25.60 \pm   0.26 $  &  $ 36.4 $  &  $ 0.07 $  &  $  20 $  &  &  $  7.6\pm 3.6\times 10^{10} $  &  &  $ <  20 $  & G &  &  $ 0.04  \pm 0.01  $  &  $  51.8  \pm 0.5 $  & T    \\
NGC1549 &  $  19.7 \pm  \phantom{1}  1.6 $  &  &  $ -24.69 \pm   0.18 $  &  $ 29.0 $  &  $ 0.10 $  &  $ 143 $  &  &  $  3.4\pm 0.7\times 10^{10} $  &  &  $  60 $  & G &  &  $ 0.13  \pm 0.01  $  &  $ 119.2  \pm 2.5 $  & T    \\
NGC1553 &  $  18.5 \pm  \phantom{1}  1.5 $  &  &  $ -25.06 \pm   0.17 $  &  $ 33.9 $  &  $ 0.34 $  &  $ 158 $  &  &  $  4.4\pm 1.1\times 10^{10} $  &  &  $ 188 $  & H &  &  \nodata  &  \nodata  &  \nodata   \\
NGC1600 &  $  66.0 \pm  \phantom{1}  9.9 $  &  &  $ -26.06 \pm   0.33 $  &  $ 24.8 $  &  $ 0.28 $  &  $   5 $  &  &  $  1.3\pm 0.5\times 10^{11} $  &  &  $ <  10 $  & D &  &  $ 0.34 \pm 0.01   $  &  $   7.4  \pm 0.6 $  &  V   \\
NGC1700 &  $  54.4 \pm  \phantom{1}  8.2 $  &  &  $ -25.59 \pm   0.33 $  &  $ 15.9 $  &  $ 0.30 $  &  $  90 $  &  &  $  8.2\pm 3.2\times 10^{10} $  &  &  $  93 $  & D &  &  $ 0.26 \pm 0.01   $  &  $  88.1  \pm 0.5 $  & T    \\
NGC2434 &  $  21.6 \pm  \phantom{1}  2.9 $  &  &  $ -23.78 \pm   0.29 $  &  $ 19.3 $  &  $ 0.08 $  &  $ 145 $  &  &  $  2.2\pm 0.6\times 10^{10} $  &  &  $  10 $  & I &  &  \nodata  &  \nodata  &  \nodata   \\
NGC2865 &  $  37.8 \pm  \phantom{1}  3.5 $  &  &  $ -24.43 \pm   0.20 $  &  $ 14.8 $  &  $ 0.22 $  &  $ 155 $  &  &  $  3.4\pm 1.0\times 10^{10} $  &  &  $  85 $  & J &  &  \nodata  &  \nodata  &  \nodata   \\
NGC3115 &  $   9.7 \pm  \phantom{1}  0.4 $  &  &  $ -24.05 \pm   0.09 $  &  $ 36.4 $  &  $ 0.57 $  &  $  42 $  &  &  $  1.6\pm 0.9\times 10^{10} $  &  &  $ 254 $  & K &  &  \nodata  &  \nodata  &  \nodata   \\
NGC3377 &  $  11.2 \pm  \phantom{1}  0.5 $  &  &  $ -22.81 \pm   0.09 $  &  $ 27.7 $  &  $ 0.40 $  &  $  42 $  &  &  $  8.0\pm 1.5\times 10^{9\phantom{0}} $  &  &  $  89 $  & D &  &  $ 0.50 \pm 0.01   $  &  $  41.0  \pm 0.2 $  &  W   \\
NGC3379 &  $  10.6 \pm  \phantom{1}  0.5 $  &  &  $ -23.85 \pm   0.11 $  &  $ 29.9 $  &  $ 0.09 $  &  $  70 $  &  &  $  1.6\pm 0.3\times 10^{10} $  &  &  $  80 $  & D &  &  $ 0.11 \pm 0.01   $  &  $  70.6   \pm 1.2 $  &  W   \\
NGC3585 &  $  20.0 \pm  \phantom{1}  1.7 $  &  &  $ -24.81 \pm   0.18 $  &  $ 32.3 $  &  $ 0.33 $  &  $ 107 $  &  &  $  3.8\pm 0.9\times 10^{10} $  &  &  $  90 $  & K &  &  \nodata  &  \nodata  &  \nodata   \\
NGC3923 &  $  22.9 \pm  \phantom{1}  3.0 $  &  &  $ -25.30 \pm   0.28 $  &  $ 43.8 $  &  $ 0.28 $  &  $  47 $  &  &  $  5.8\pm 1.6\times 10^{10} $  &  &  $ < 100 $  & L &  &  $ 0.36  \pm 0.01  $  &  $  47.9  \pm 0.5 $  &  U   \\
NGC4125 &  $  23.9 \pm  \phantom{1}  2.7 $  &  &  $ -25.03 \pm   0.25 $  &  $ 33.0 $  &  $ 0.37 $  &  $  80 $  &  &  $  5.5\pm 1.4\times 10^{10} $  &  &  $ 185 $  & D &  &  $ 0.45 \pm 0.01   $  &  $  82.0   \pm 0.1 $  & T    \\
NGC4261 &  $  31.6 \pm  \phantom{1}  2.8 $  &  &  $ -25.24 \pm   0.19 $  &  $ 25.5 $  &  $ 0.16 $  &  $ 158 $  &  &  $  5.1\pm 1.2\times 10^{10} $  &  &  $  25 $  & M &  &  $ 0.21 \pm 0.01   $  &  $ 158.4   \pm  0.5 $  & V    \\
NGC4365 &  $  20.4 \pm  \phantom{1}  1.6 $  &  &  $ -24.91 \pm   0.17 $  &  $ 40.7 $  &  $ 0.24 $  &  $  37 $  &  &  $  4.3\pm 1.2\times 10^{10} $  &  &  $   7 $  & N &  &  $ 0.24 \pm 0.01   $  &  $  41.6  \pm 0.3 $  & T    \\
NGC4374 &  $  18.4 \pm  \phantom{1}  0.9 $  &  &  $ -25.10 \pm   0.11 $  &  $ 34.8 $  &  $ 0.06 $  &  $ 148 $  &  &  $  5.7\pm 1.0\times 10^{10} $  &  &  $  30 $  & M &  &  $ 0.13 \pm 0.01   $  &  $ 125.7  \pm 0.3 $  & W    \\
NGC4406 &  $  17.1 \pm  \phantom{1}  1.1 $  &  &  $ -25.07 \pm   0.14 $  &  $ 59.7 $  &  $ 0.25 $  &  $ 120 $  &  &  $  5.9\pm 1.3\times 10^{10} $  &  &  $ < 100 $  & D &  &  $ 0.24 \pm 0.01   $  &  $ 123.2  \pm 0.5 $  & V    \\
NGC4472 &  $  16.3 \pm  \phantom{1}  0.8 $  &  &  $ -25.66 \pm   0.10 $  &  $ 59.2 $  &  $ 0.15 $  &  $ 153 $  &  &  $  8.8\pm 1.7\times 10^{10} $  &  &  $ 117 $  & D &  &  $ 0.17 \pm 0.01   $  &  $ 160.6  \pm 0.5 $  &  V   \\
NGC4494 &  $  17.1 \pm  \phantom{1}  0.9 $  &  &  $ -24.16 \pm   0.11 $  &  $ 30.8 $  &  $ 0.13 $  &  $ 170 $  &  &  $  2.7\pm 0.4\times 10^{10} $  &  &  $  73 $  & D &  &  $ 0.18 \pm 0.01   $  &  $ 177.7  \pm 0.4 $  &  T   \\
NGC4526 &  $  16.9 \pm  \phantom{1}  1.6 $  &  &  $ -24.67 \pm   0.20 $  &  $ 43.8 $  &  $ 0.63 $  &  $ 111 $  &  &  $  2.8\pm 0.7\times 10^{10} $  &  &  $ 205 $  & L &  &  \nodata  &  \nodata  &  \nodata   \\
NGC4552 &  $  15.3 \pm  \phantom{1}  1.0 $  &  &  $ -24.20 \pm   0.14 $  &  $ 25.4 $  &  $ 0.08 $  &  $ 150 $  &  &  $  2.3\pm 0.4\times 10^{10} $  &  &  $  40 $  & O &  &  $ 0.05 \pm 0.01   $  &  $ 122.7  \pm 1.5 $  & W    \\
NGC4555 &  $  97.4 \pm  14.6 $  &  &  $ -25.78 \pm   0.33 $  &  $ 10.9 $  &  $ 0.16 $  &  $ 125 $  &  &  $  5.1\pm 1.7\times 10^{10} $  &  &  \nodata  &  \nodata  &  &  \nodata  &  \nodata  &  \nodata   \\
NGC4564 &  $  15.0 \pm  \phantom{1}  1.2 $  &  &  $ -22.94 \pm   0.17 $  &  $ 19.9 $  &  $ 0.54 $  &  $  50 $  &  &  $  6.8\pm 1.5\times 10^{9\phantom{0}} $  &  &  $ 155 $  & D &  &  $ 0.53 \pm 0.02  $  &  $  48.1  \pm 0.3 $  & W    \\
NGC4621 &  $  18.3 \pm  \phantom{1}  1.7 $  &  &  $ -24.56 \pm   0.20 $  &  $ 32.9 $  &  $ 0.37 $  &  $ 165 $  &  &  $  2.9\pm 1.0\times 10^{10} $  &  &  $ 140 $  & D &  &  $ 0.35 \pm 0.01   $  &  $ 162.9  \pm 0.1 $  & W    \\
NGC4636 &  $  14.7 \pm  \phantom{1}  0.9 $  &  &  $ -24.41 \pm   0.13 $  &  $ 59.3 $  &  $ 0.22 $  &  $ 148 $  &  &  $  2.7\pm 0.6\times 10^{10} $  &  &  $  30 $  & M &  &  $ 0.17 \pm 0.01   $  &  $ 147.1  \pm 0.6 $  & V    \\
NGC4649 &  $  16.8 \pm  \phantom{1}  1.2 $  &  &  $ -25.39 \pm   0.15 $  &  $ 45.2 $  &  $ 0.19 $  &  $ 110 $  &  &  $  5.9\pm 1.0\times 10^{10} $  &  &  $ 110 $  & D &  &  $ 0.18 \pm 0.01   $  &  $ 104.0  \pm 0.4 $  & V    \\
NGC4697 &  $  11.7 \pm  \phantom{1}  0.8 $  &  &  $ -23.98 \pm   0.14 $  &  $ 42.4 $  &  $ 0.33 $  &  $  65 $  &  &  $  1.7\pm 0.9\times 10^{10} $  &  &  $ 110 $  & P &  &  $ 0.45 \pm 0.01   $  &  $  65.7  \pm 0.4 $  & T    \\
NGC5018 &  $  39.9 \pm  \phantom{1}  6.0 $  &  &  $ -25.27 \pm   0.33 $  &  $ 15.6 $  &  $ 0.28 $  &  $  95 $  &  &  $  6.3\pm 3.4\times 10^{10} $  &  &  $  65 $  & Q &  &  $ 0.33 \pm 0.01   $  &  $  93.8  \pm 0.5 $  & T    \\
NGC5044 &  $  31.2 \pm  \phantom{1}  4.0 $  &  &  $ -24.76 \pm   0.28 $  &  $ 25.3 $  &  $ 0.04 $  &  $  20 $  &  &  $  4.0\pm 1.8\times 10^{10} $  &  &  $  70 $  & R &  &  $ 0.07 \pm 0.01   $  &  $  3.1  \pm 5.4 $  & T    \\
NGC5102 &  $   4.0 \pm  \phantom{1}  0.2 $  &  &  $ -21.09 \pm   0.14 $  &  $ 79.3 $  &  $ 0.50 $  &  $  50 $  &  &  $  2.4\pm 0.4\times 10^{9\phantom{0}} $  &  &  \nodata  &  \nodata  &  &  \nodata  &  \nodata  &  \nodata   \\
NGC5171 &  $  99.3 \pm  14.9 $  &  &  $ -24.95 \pm   0.33 $  &  $ 10.8 $  &  $ 0.22 $  &  $ 175 $  &  &  $  6.2\pm 2.0\times 10^{10} $  &  &  \nodata  &  \nodata  &  &  \nodata  &  \nodata  &  \nodata   \\
NGC5532 &  $ 104.5 \pm  15.7 $  &  &  $ -26.33 \pm   0.33 $  &  $ 16.0 $  &  $ 0.24 $  &  $ 155 $  &  &  $  1.2\pm 0.5\times 10^{11} $  &  &  \nodata  &  \nodata  &  &  \nodata  &  \nodata  &  \nodata   \\
NGC5845 &  $  25.9 \pm  \phantom{1}  2.5 $  &  &  $ -22.96 \pm   0.21 $  &  $  4.9 $  &  $ 0.32 $  &  $ 140 $  &  &  $  5.3\pm 1.4\times 10^{9\phantom{0}} $  &  &  $ 110 $  & S &  &  $ 0.28 \pm 0.01   $  &  $ 143.1  \pm 0.3 $  & W    \\
NGC5846 &  $  24.9 \pm  \phantom{1}  2.3 $  &  &  $ -25.04 \pm   0.20 $  &  $ 34.5 $  &  $ 0.06 $  &  $  30 $  &  &  $  4.4\pm 1.0\times 10^{10} $  &  &  $ <  10 $  & R &  &  $ 0.05 \pm 0.01   $  &  $  64.0  \pm 1.9 $  & W    \\
NGC6482 &  $  58.9 \pm  \phantom{1}  8.8 $  &  &  $ -25.48 \pm   0.33 $  &  $ 12.6 $  &  $ 0.28 $  &  $  65 $  &  &  $  1.1\pm 0.4\times 10^{11} $  &  &  \nodata  &  \nodata  &  &  $ 0.23 \pm 0.01   $  &  $  66.8  \pm 0.3 $  & T    \\
NGC7052 &  $  70.2 \pm  10.5 $  &  &  $ -25.66 \pm   0.33 $  &  $ 21.8 $  &  $ 0.50 $  &  $  60 $  &  &  $  4.0\pm 1.2\times 10^{10} $  &  &  \nodata  &  \nodata  &  &  $ 0.46 \pm 0.01   $  &  $  63.5  \pm 0.7 $  & U    \\
NGC7618 &  $  77.2 \pm  11.6 $  &  &  $ -25.40 \pm   0.33 $  &  $ 11.6 $  &  $ 0.20 $  &  $   5 $  &  &  $  4.6\pm 2.2\times 10^{10} $  &  &  \nodata  &  \nodata  &  &  \nodata  &  \nodata  & \nodata    \\

\enddata
\tablenotetext{a}{Distances and associated errors in Mpc, 
taken from \citet{Tonry} and LEDA \citep{LEDA}, LEDA errors 
are assumed to be $15\%$.}
\tablenotetext{b}{Data taken from the 2MASS extended source catalog 
\citep{2MASS}: absolute magnitude ($K$-band), half-light radius in 
$\arcsec$, effective ellipticity and position angle (all $J$-band).}
\tablenotetext{c}{Absolute blue luminosity in units of 
$L_{B,\Sun}=5.2\times 10^{32} \erg\s^{-1}$}
\tablenotetext{d}{Rotational velocities in $\rm km\, s^{-1}$ 
evaluated between $0.6-0.9\,R_{\rm J}$, taken from Hyperleda 
\citep{Hyperleda} references. 
A: \citet{SP97a}, 
B: \citet{PS94}, 
C: \citet{Bon+95}, 
D: \citet{BSG94}, 
E: \citet{Don+95}, 
F: \citet{GCB98}, 
G: \citet{Lon+94}, 
H: \citet{LRB98}, 
I: \citet{CD94a}, 
J: \citet{Bet92}, 
K: \citet{F97}, 
L: \citet{PHC97}, 
M: \citet{DB88}, 
N: \citet{SP98a}, 
O: \citet{SP97b}, 
P: \citet{BDI90}, 
Q: \citet{CD94b}, 
R: \citet{CDB-93}, 
S: \citet{SP02_2}.
}
\tablenotetext{e}{Ellipticity and position angle, evaluated 
between $0.6-0.9\,R_{\rm J}$ from published optical surface 
photometry. T: \citet{GoudfrooijPhot}, U: \citet{BenderPhot},
V: \citet{PeletierPhot}, W: \citet{MDMPhot}.}

\end{deluxetable*}


\section{Results}\label{s3.results}

\subsection{X-ray vs. Optical Luminosity}

\newcommand{\mywidth}{0.\textwidth}
\begin{figure*}
\includegraphics[width=\textwidth]{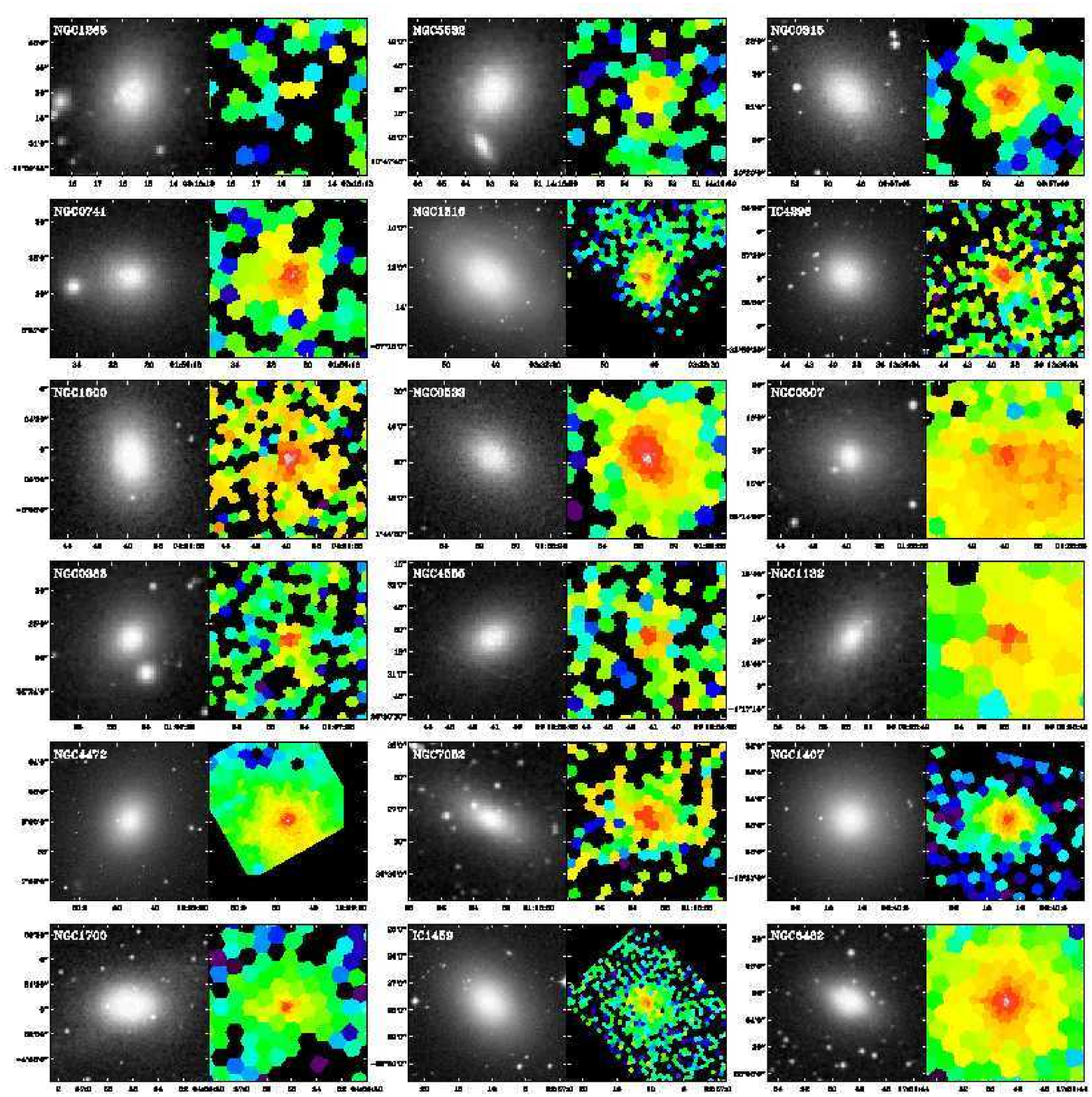}
\caption{Adaptively binned {\it Chandra} X-ray gas surface brightness
maps (right) and optical DSS $R$-band images (left). The objects are
ordered by 2MASS $K$ band luminosity, starting with the most luminous
galaxy on the top left of the page, and decreasing to the right and
then to the next row. The physical scale of each image is
$50\kpc\times 50\kpc$, except where indicated by an individually
attached scale bar. The color range of the X-ray gas distribution is
scaled logarithmically between $5\times 10^{-11}$ and $3\times
10^{-7}\,{\rm photons \, sec^{-1} cm^{-2} \, arcsec^{-2}}$. The x and
y-axes are labelled according to right ascension and declination
(2000), respectively. \label{f.chandradss1}}
\end{figure*}
\begin{figure*}
\setcounter{figure}{1}
\includegraphics[width=\textwidth]{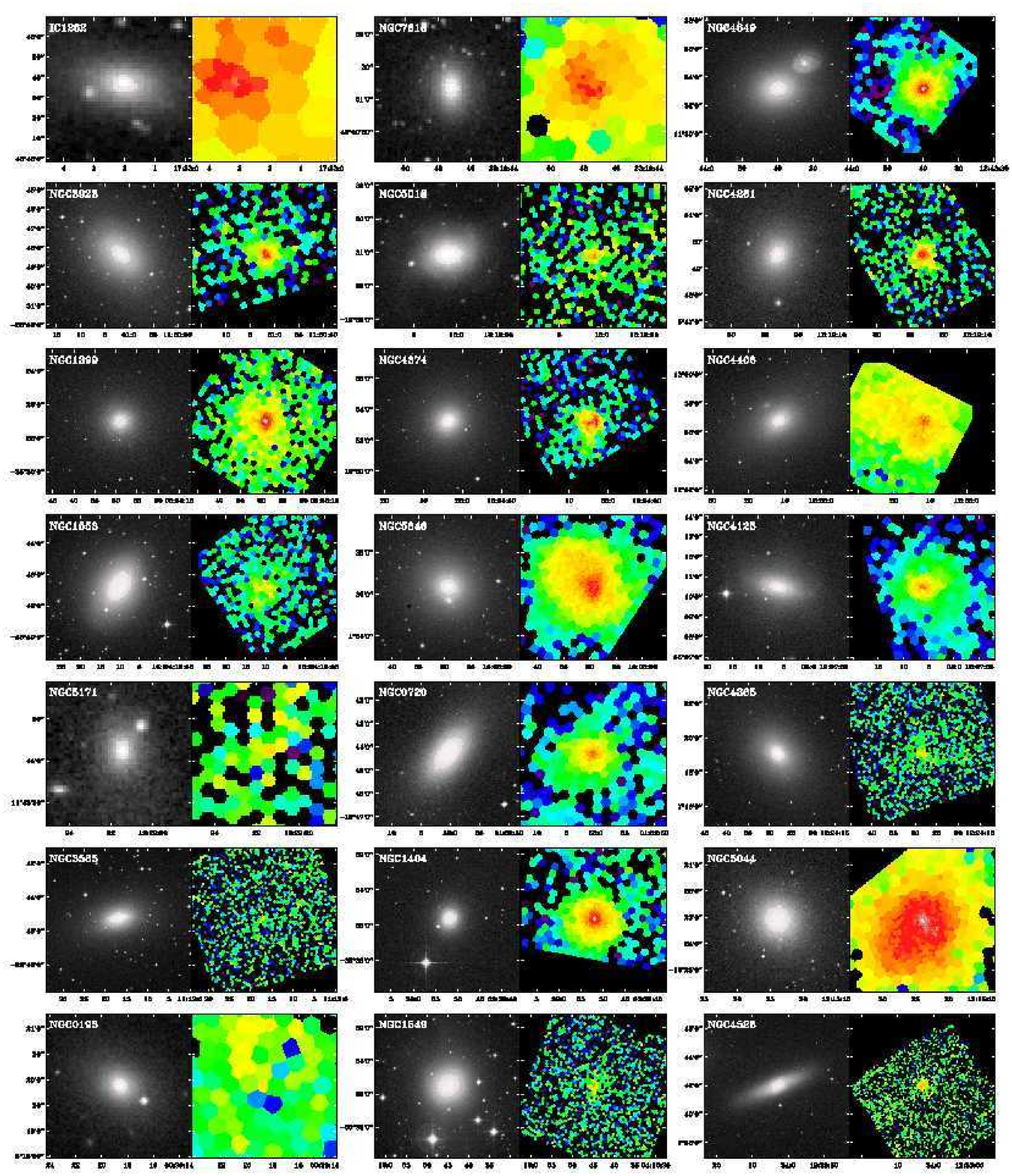}
\caption[]{Continued \label{f.chandradss2}}
\end{figure*}
\begin{figure*}
\setcounter{figure}{1}
\includegraphics[width=\textwidth]{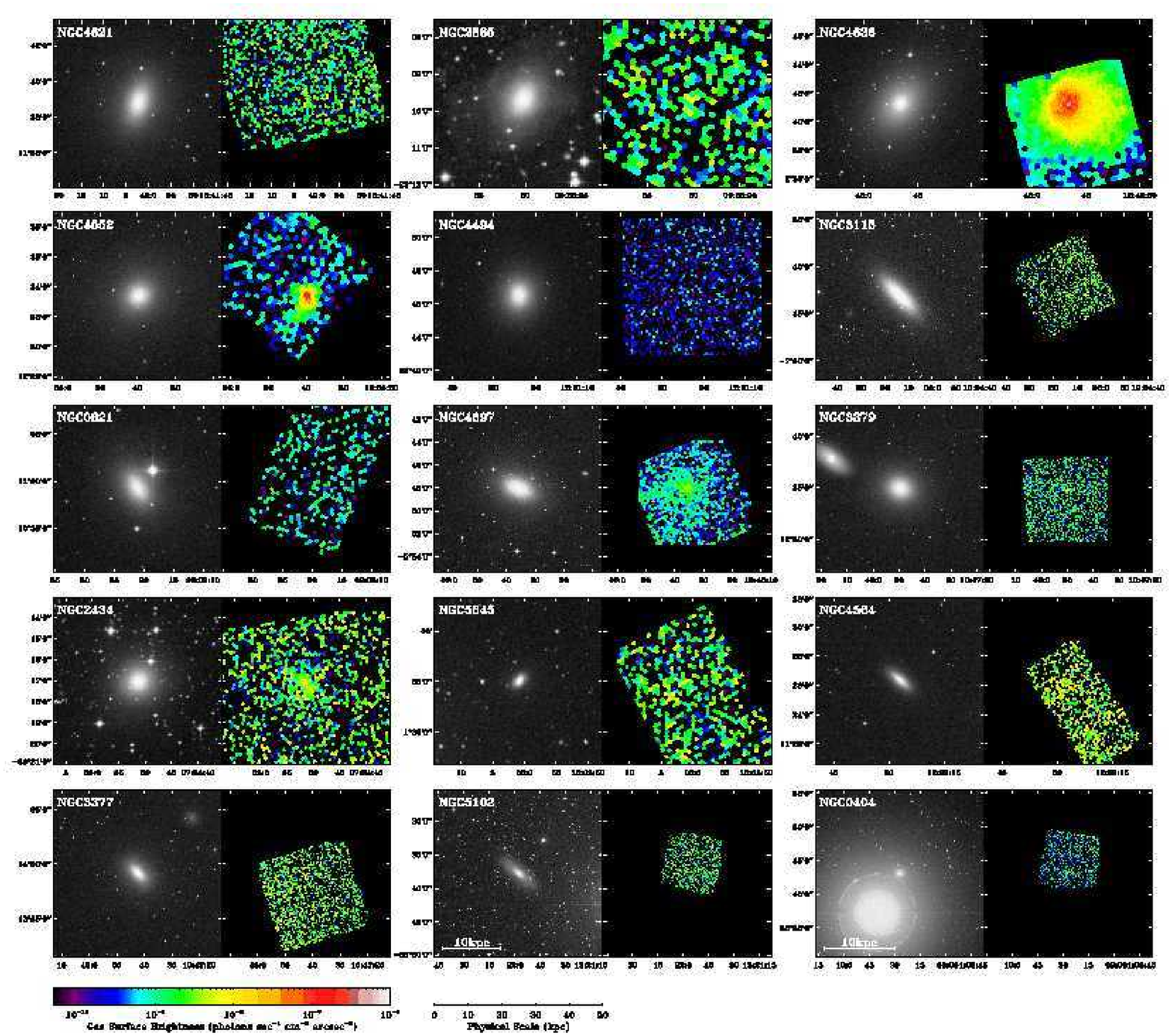}
\caption[]{Continued \label{f.chandradss3}}
\end{figure*}

\begin{figure}
\plotone{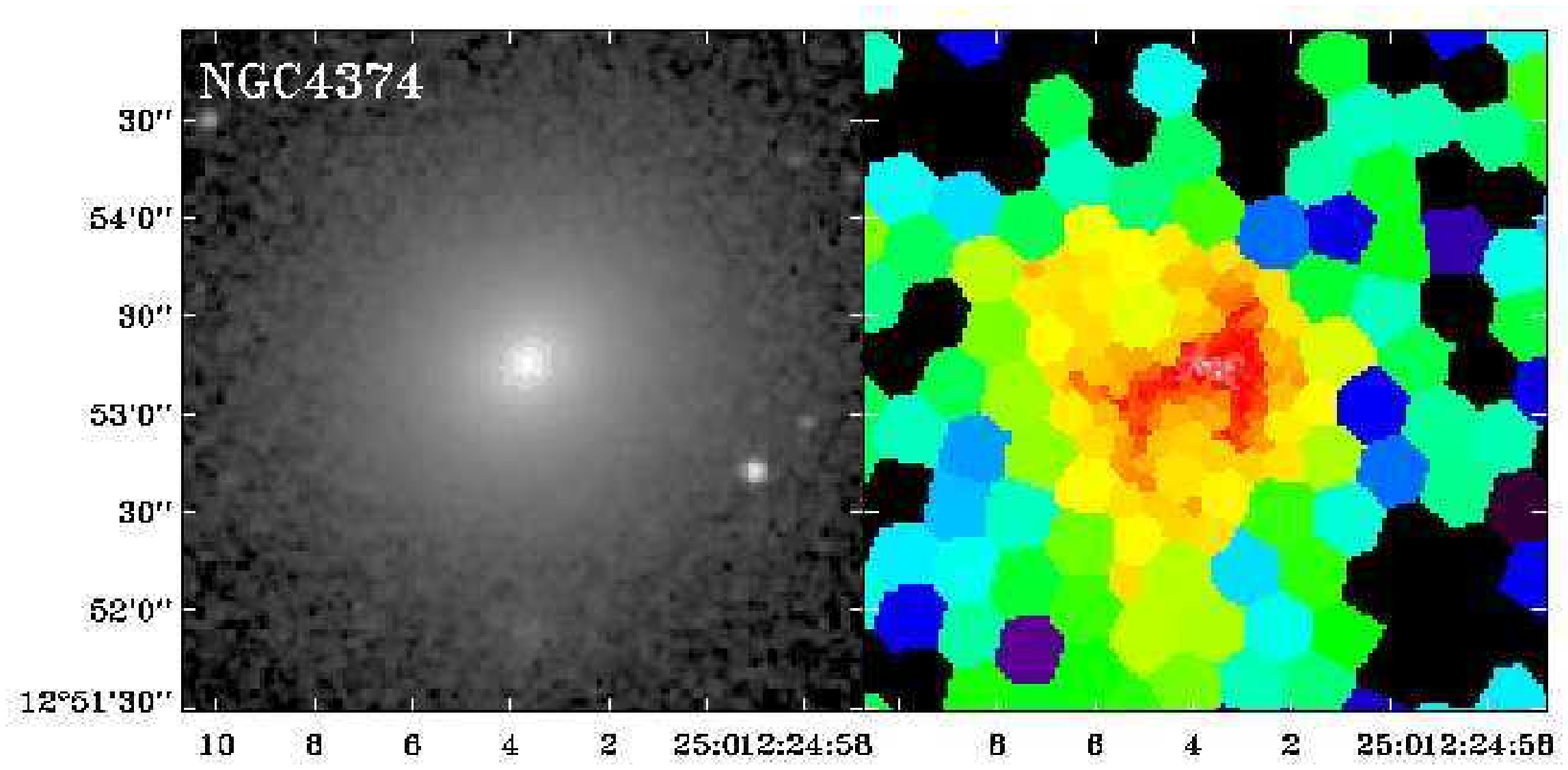}
\caption{Adaptively binned {\it Chandra} X-ray gas surface brightness maps (right) and combined 2MASS $J$, $H$ and $K_s$-band images (left). Figures \ref{f.chandra2mass}a-ag showing all 33 galaxies with sufficient data to compute the X-ray ellipticity $\epsilon_X$ are available in high-resolution in the electronic edition of the Journal. The printed edition contains only a sample (NGC~4374). When available, 2MASS images were taken from the 2MASS Large Galaxy Atlas \citep[][\texttt{http://irsa.ipac.caltech.edu/applications/2MASS/LGA/atlas.html}]{2MASSLGA}, otherwise images are taken from the 2MASS Extended Source Catalog \citep{2MASS}. The physical scale of each image is $6 R_{\rm J}$ across. The color range of the X-ray gas distribution is scaled logarithmically between $5\times 10^{-11}$ and $1\times 10^{-6}\,{\rm photons \, sec^{-1} cm^{-2} \, arcsec^{-2}}$, and is identical to that of Figure \ref{f.chandradss1}. The x and y-axes are labelled according to right ascension and declination (2000), respectively. 
\label{f.chandra2mass}}
\end{figure}

For the first time, we have spatially isolated the hot gas emission in
normal elliptical galaxies to allow an unbiased morphological analysis
of a large sample of objects. Figure \ref{f.chandradss1} shows the
X-ray gas gallery, along with the optical DSS-2 $R$-band images for
each object. We rescale all images to show the same physical size of
$50\kpc$, except for the two closest galaxies, where we indicate the
size by a separate scale bar. We have ordered Figure
\ref{f.chandradss1} with decreasing absolute $K$ luminosity, a good
indicator of the total stellar mass of the host galaxies. This is an intuitive
way to visualize the large scatter in the well-known 
$L_{\rm X}\propto L_{\rm B}^2$ relation \citep[and references 
therein]{OSullivanLxLb}, one of the elementary empirical
correlations connecting X-ray and optical parameters. Its scatter shows up as a 
large spread in X-ray properties for galaxies that have
similar stellar contents, i.e. that are close to each other in the
figure. Galaxies with similar optical properties can have gas
luminosities differing by orders of magnitude (e.g. NGC~5044 and
NGC~1549, NGC~4555 and NGC~0533). At the same time, galaxies with
comparable gas contents can have extremely different optical
appearances (e.g. NGC~1404 and NGC~4649).

\begin{figure}
\plotone{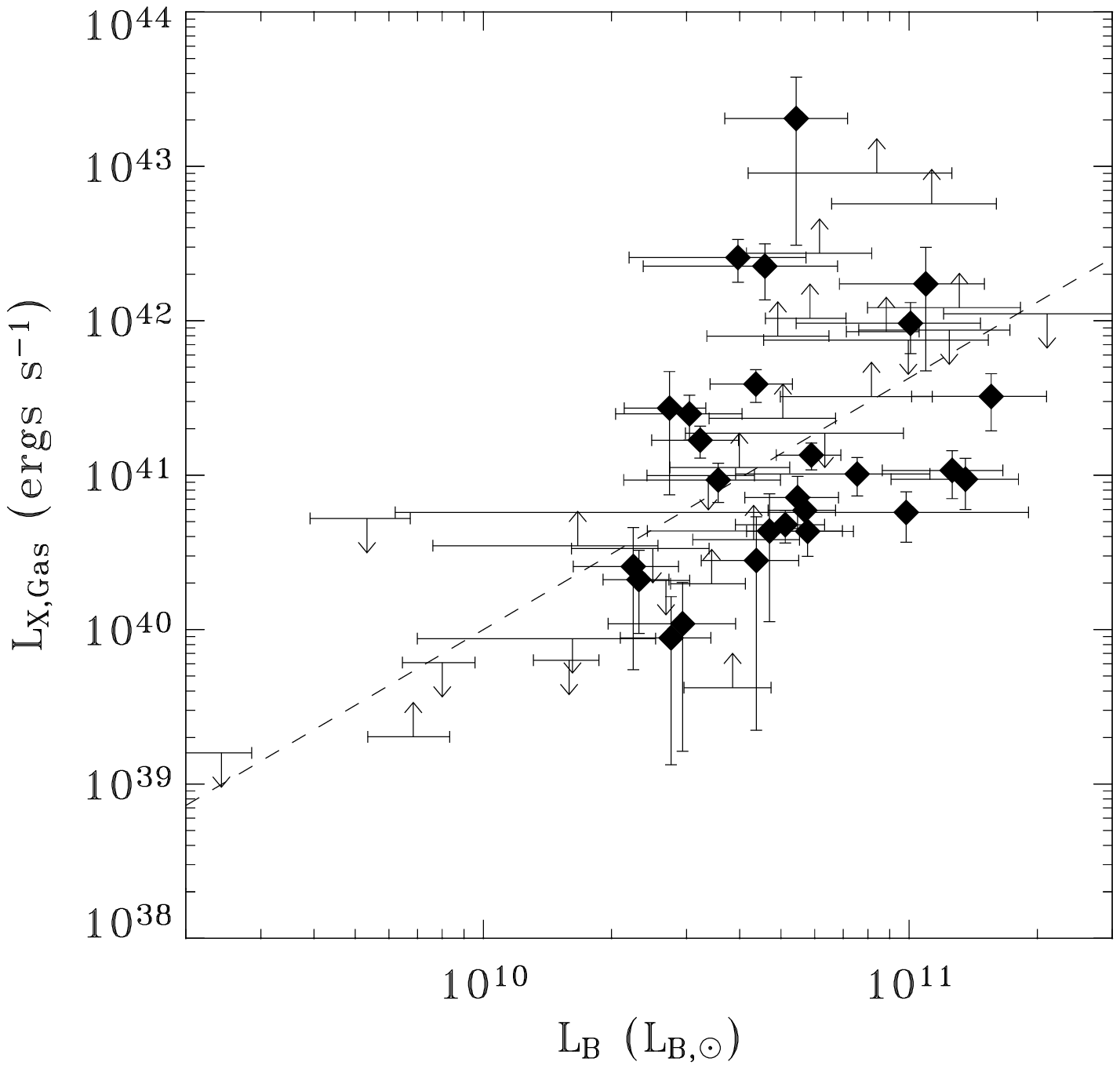}
\caption{Total X-ray gas luminosity in $\erg\s^{-1}$ as a function of
absolute blue luminosity in units of blue solar luminosity
($L_{B,\Sun}=5.2\times 10^{32} \erg\s^{-1}$) taken from LEDA. The
dashed line shows the best fit by \citet{OSullivanLxLb}, which they
statistically correct for the expected contribution from unresolved
point sources. Our data are consistent with their
relation.\label{f.lxlb}}
\end{figure}

\citet{OSullivanLxLb} also observe a flattening of the $L_{\rm X}$--$L_{\rm B}$ 
relation toward lower blue luminosities. They attribute this trend to
the increasing relative importance of point source emission, which is
unresolved in their ROSAT and Einstein data. Our $L_{\rm
X,Gas}$--$L_{\rm B}$ relation, with the stellar source contribution
removed (Figure \ref{f.lxlb}), is consistent with a single slope, and
agrees well with \cite{OSullivanLxLb}'s best fit (dashed
line). Nonetheless, we still see a large scatter in this relation,
covering almost two orders of magnitude in X-ray luminosity for a
given absolute $B$ magnitude. A recent study by \citet{EllisLxLb}
shows a similar relation between X-ray and $K$-band luminosities, with
an almost identical scatter, suggesting a large scatter in X-ray
luminosities as the dominant cause.

\subsection{X-ray vs. Optical Morphology}

Not only do the luminosities of the hot gas and stellar components
differ vastly, there are also significant differences between the
X-ray and optical morphology. Figure \ref{f.chandradss1}
shows that it is impossible to predict the gas morphology from looking
at the stellar distribution, or vice versa. There are some optically
flat galaxies with round X-ray isophotes (e.g. NGC~0720), and others
with very flat X-ray emission (e.g. NGC~1700). Optically round
galaxies show a similar range in X-ray ellipticities, ranging from
round (e.g. NGC~1404) to flat (e.g. NGC~0533).

Many galaxies, in fact, appear asymmetrically disturbed. We discuss
this asymmetry and its origin in Paper II. For now, it is important to
first quantify the overall shape of the gas emission. If the gas is in
true hydrostatic equilibrium, the hot gas isophotes should very nearly trace the
projected equipotentials \citep[e.g.][]{BinneyGalDynamics,
BuoteGeomtest}. Thus, we construct X-ray gas ellipticity and position
angle profiles, shown in Figure \ref{f.epsprofiles1} (solid circles),
with published optical profiles overlaid. The optical profiles are
generally ``well-behaved'' and reveal only modest radial trends in
their ellipticities and position angles. Optical isophotal twists, if
present at all, rarely exceed a few degrees.

The X-ray profiles tell a completely different story. Often, the gas
ellipticities change rapidly as a function of radius (e.g. NGC~1399,
NGC~4374). These features are often accompanied by sudden changes 
in
position angle (e.g. NGC~4636, NGC~5044). Isophotal twists in the gas
emission are common, and only very few objects show profiles that are
consistent with a single major axis orientation (e.g. NGC~3923,
NGC~1700). In many cases isophotal twists or sudden changes in
ellipticity also coincide with asymmetries in the gas images
(e.g. NGC~5044).

\begin{figure*}
\includegraphics[width=\textwidth]{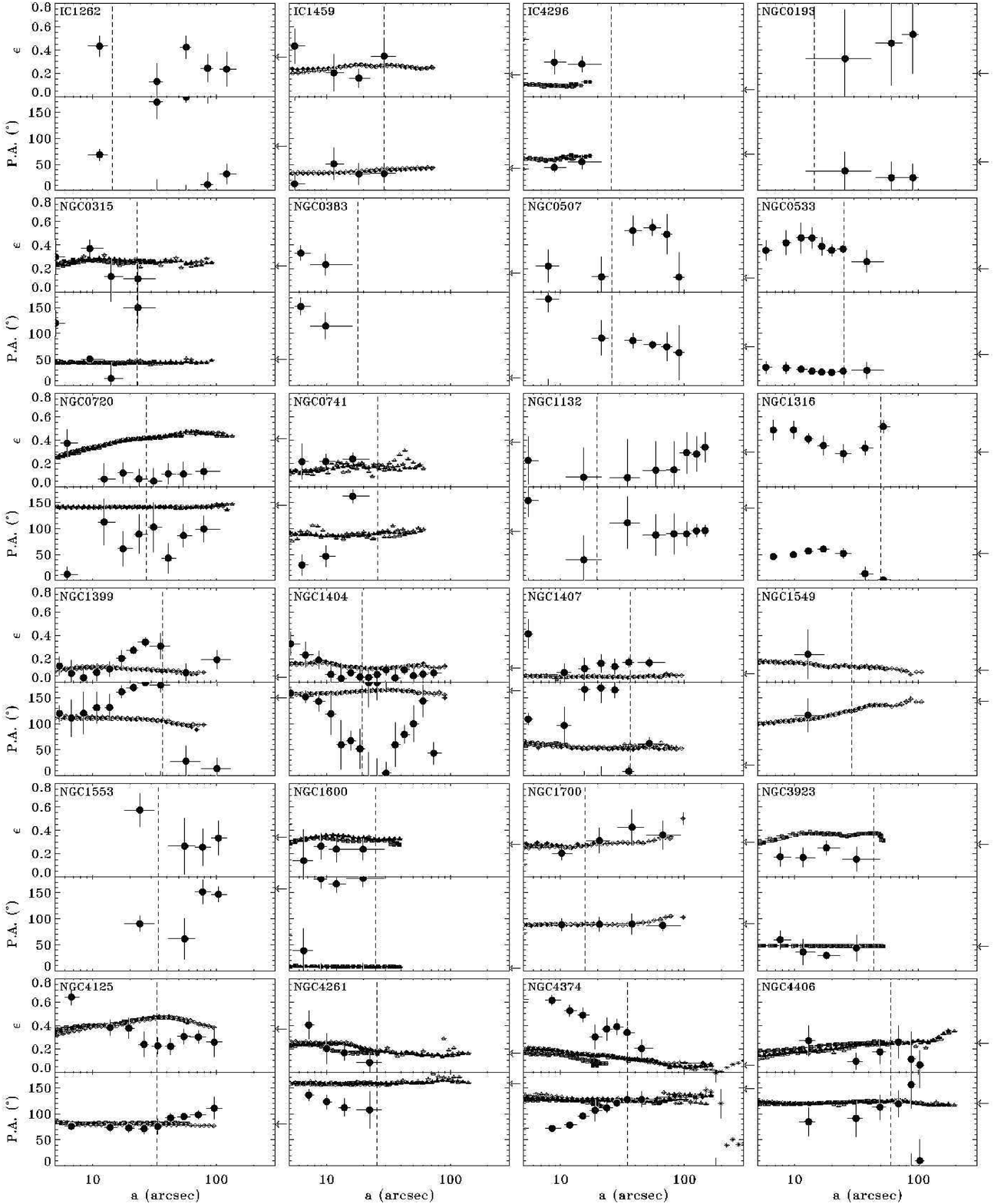}
\caption{Radial isophotal ellipticity and position angle profiles for
the subset of 36 elliptical galaxies containing sufficient
signal. Large filled circles with error bars denote X-ray gas
profiles; other symbols denote optical profiles as indicated in the
figure. The 2MASS ellipticity and position angle are marked with
arrows at the right border. The vertical dashed lines mark the 2MASS
$J$-band effective radius. \label{f.epsprofiles1}}
\end{figure*}

\begin{figure*}
\setcounter{figure}{4}
\includegraphics[width=\textwidth]{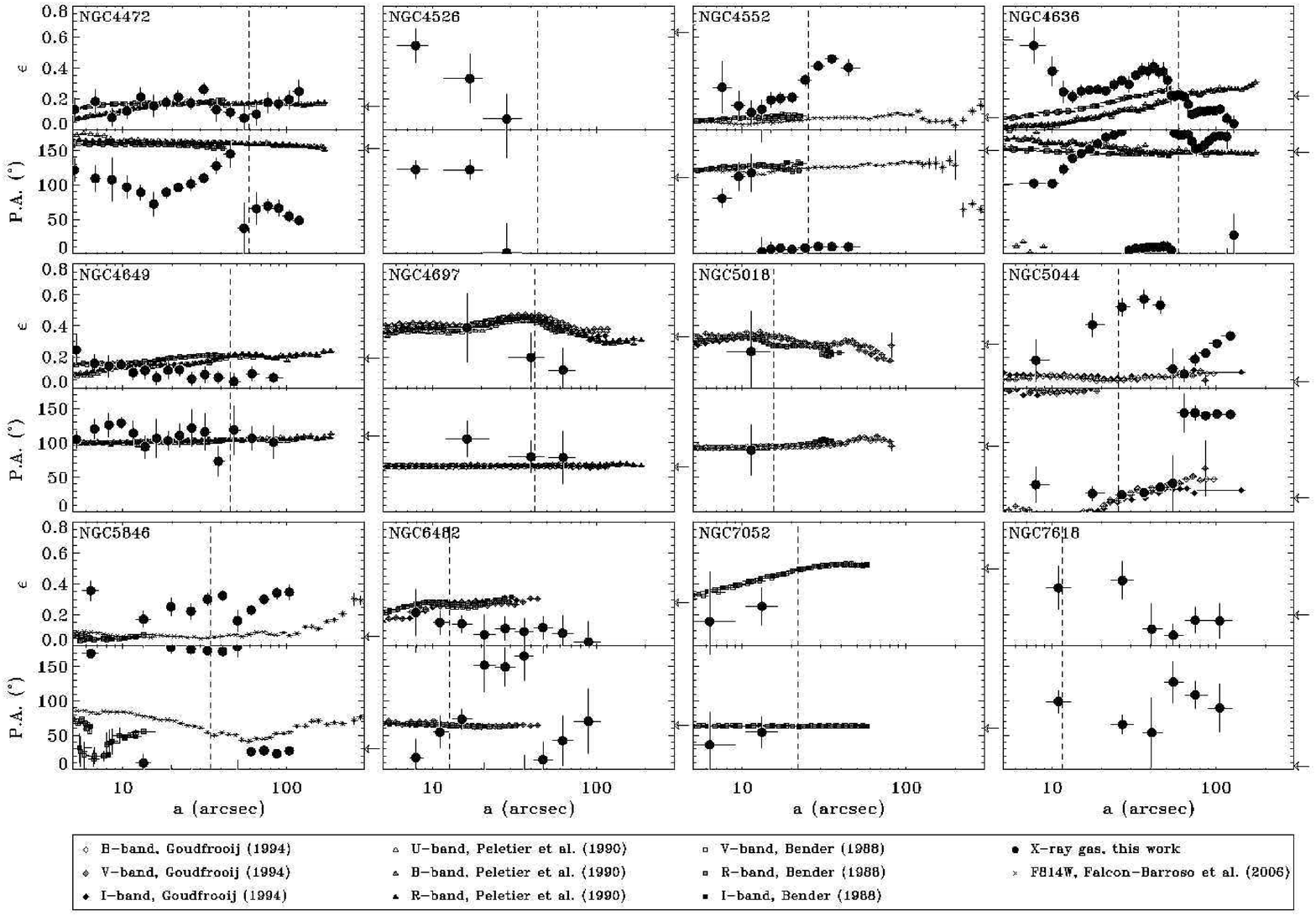}
\caption[]{Continued. \label{f.epsprofiles2}}
\end{figure*}

\begin{figure}
\plotone{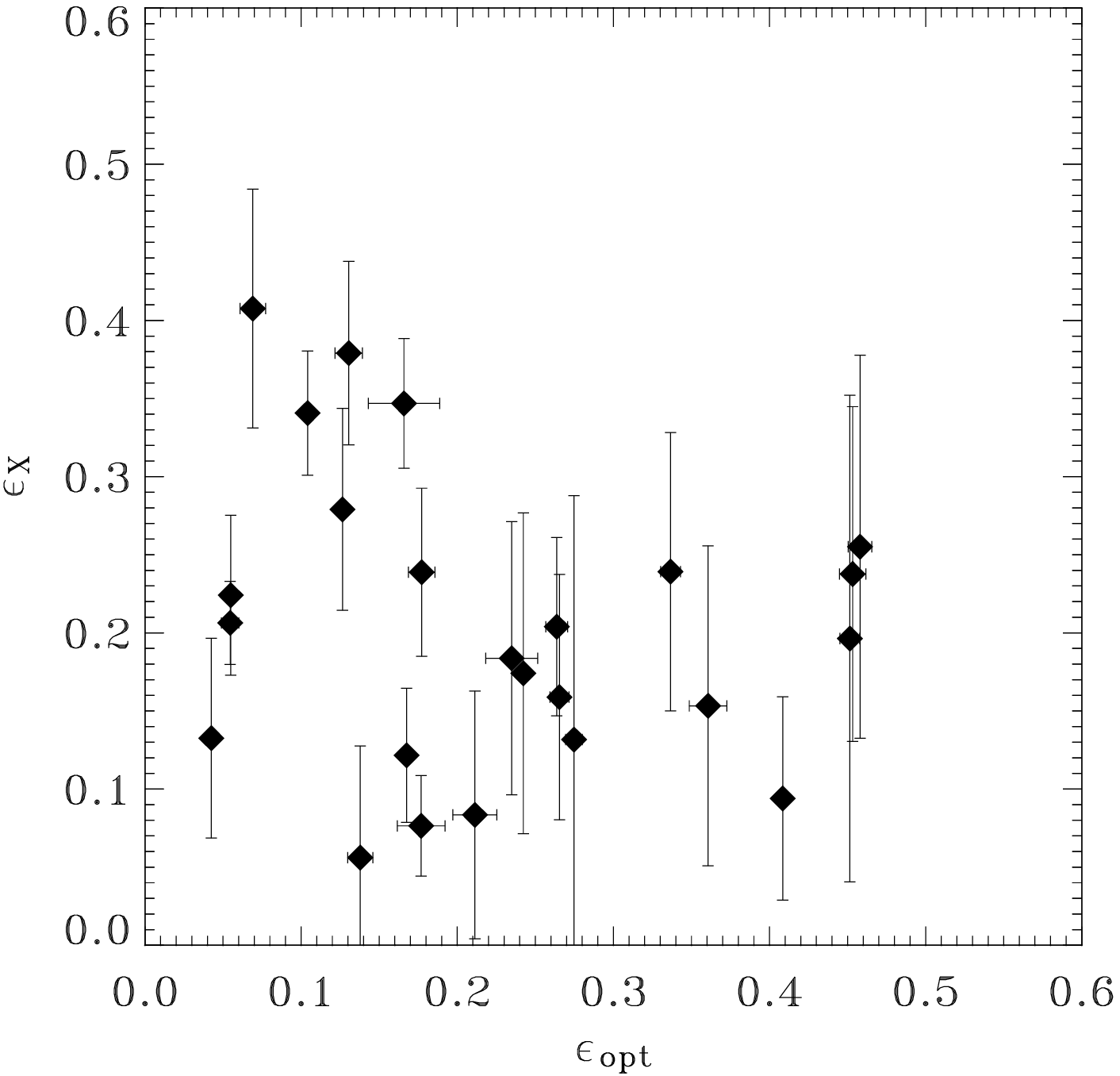}
\caption{X-ray gas ellipticity vs. optical ellipticity, both evaluated
between $0.6-0.9\,R_{\rm J}$. There is no correlation, contrary to
what what would be expected if the gas were in hydrostatic equilibrium
in a stellar-mass-dominated potential. See also Figure \ref
{f.tss_simulations} for theoretical predictions. \label{f.epsilon}}
\end{figure}

It is generally agreed that the potentials of early-type galaxies are
stellar-mass dominated inside 1 or 2 optical effective radii
\citep[e.g.][]{LoewWhite,
MamonDarkmatterI,Lintott,Humphrey}. To test whether the X-ray gas
flattening is consistent with hydrostatic equilibrium, we extract mean
ellipticities for the hot gas emission and the starlight between $0.6$ and
$0.9\,R_{\rm J}$. Figure \ref{f.chandra2mass} shows expanded views of the
central regions of the galaxies for which gas ellipticities are measurable
in this region, compared with 2MASS images. The radial range is chosen
to maximize the available number of objects with valid optical and X-ray
ellipticity profiles in the same interval, resulting in a subset of 24 galaxies.
This subset spans more than an order of magnitude in optical luminosity
and nearly two orders of magnitude in X-ray gas luminosity. Optical
ellipticities are representative of the general population,
and X-ray gas fractions range from 57\% to 100\%.
Figure \ref{f.epsilon} shows that there is absolutely no correlation between
optical and gas ellipticities. We have verified that this result holds for
other choices of extraction annuli ranging between $0.5$ and $1.2\,R_{\rm J}$,
and for different choices of optical bandpass.
The large scatter is not caused by systems with obviously disturbed
X-ray morphology. Of the 8 most highly disturbed systems (quantified by
the asymmetry index defined in Paper II), 6 do not have published
optical surface photometry, and are not plotted in Figure \ref{f.epsilon};
removing the remaining two (NGC 4374 and NGC 5044) makes an insignificant
difference. Excluding group member galaxies, or
systems with significant optical isophotal twists or ellipticity gradients
also has negligible effect.

This lack of correlation is very surprising if
hydrostatic equilibrium holds precisely. Roughly speaking, if one
assumes an oblate logarithmic potential, the equipotential surfaces
should be about one third as flattened as the underlying density
distribution \citep{BinneyGalDynamics}. Instead, we find that the
X-ray isophotes are often much flatter than the stellar isophotes.
We fully discuss the implications of this result, with detailed modeling,
in \S\ \ref{s.implhydroeq} below.

We notice a weak tendency for galaxies to have their X-ray gas major
axis aligned with the stellar distribution. Given the fact 
that X-ray isophotal twists often correlate with the orientation of
asymmetries in the gas, it is not clear whether this alignment is a
consequence of the underlying potential. An alternative is that it is
a consequence of a misalignment between radio sources optical major
axes \citep{PalimakaRadioopt}, which is present in our sample and
which we fully discuss in Paper II.

\subsection{Rotational Support}

\begin{figure}
\plotone{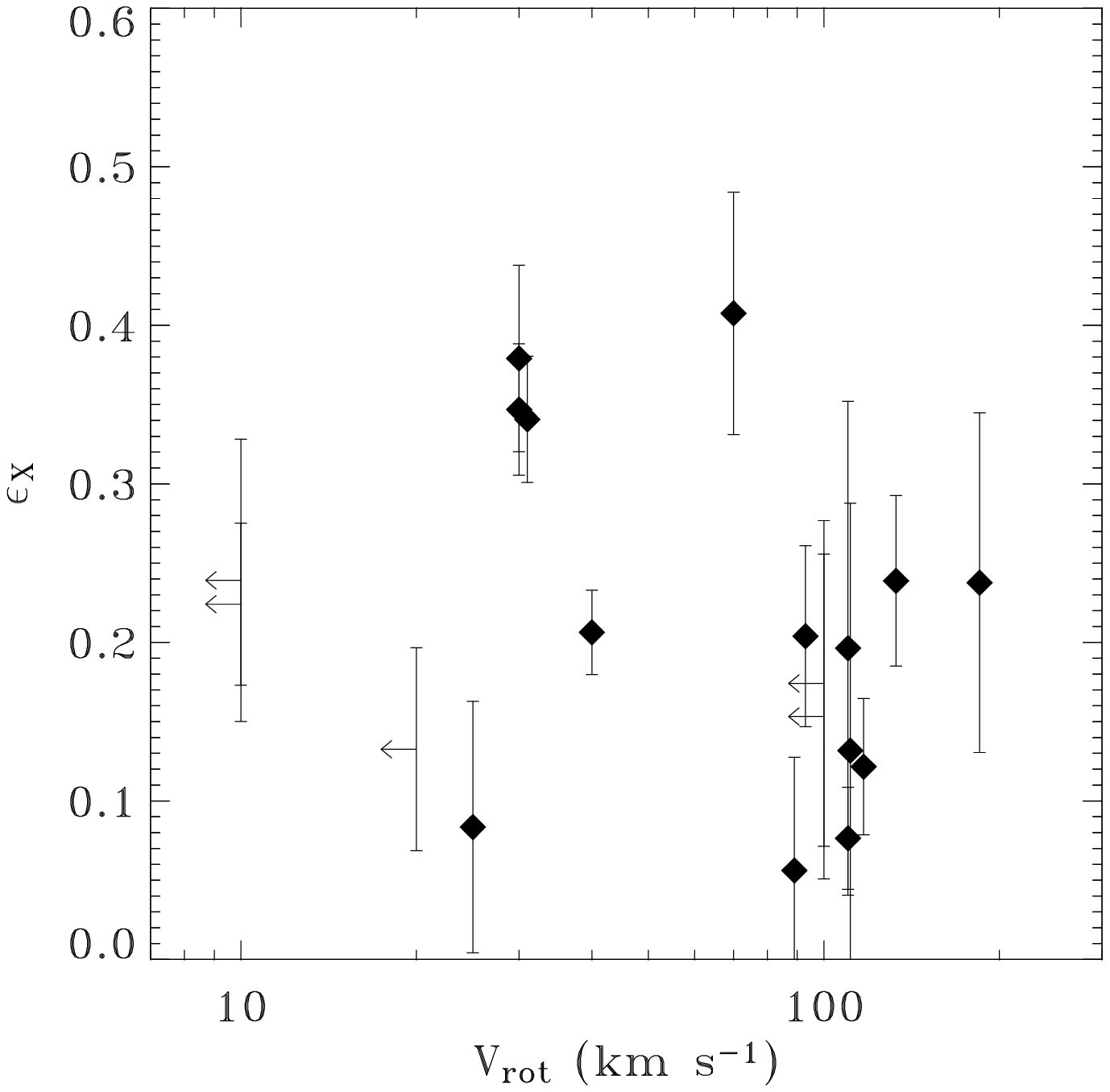}
\caption{X-ray ellipticity as a function of rotational stellar
velocity between $0.6-0.9\,R_{\rm J}$. No correlation is present,
contrary to what is expected if the hot gas would be due to stellar
mass-loss and subsequently settle into a cooling
disk. \label{f.vroteps}}
\end{figure}

A significant fraction of the hot gas in elliptical galaxies is
believed to be due to stellar mass-loss
\citep[e.g.][]{BrighentiRotating}. Thus, the gas should contain a
significant amount of specific angular momentum, causing it to settle
into a flattened rotating disk \citep{BrighentiDisks}. This scenario
predicts a relationship between the stellar rotational velocity and
the X-ray gas ellipticity, with faster rotating systems exhibiting
flatter gas isophotes. Figure \ref{f.vroteps} shows the observed X-ray
ellipticity between $0.6$ and $0.9R_{\rm J}$ as a function of mean
stellar rotational velocity in the same radial range. In agreement
with previous ROSAT and Einstein observations of elliptical galaxies
\citep{BregmanRotating}, we find no evidence for any correlation
between these parameters. A Spearman rank analysis yields a
probability of 54\% for the null hypothesis of no correlation.
While this is formally an upper bound on the probability due the presence
of finite measurement errors and the presence of upper limits, it is obvious
from the figure that the errors are not responsible for the absence of
correlation.

Such a lack of correlation could still be consistent with an
alternative scenario, where the majority of the hot gas is not due to
stellar mass loss, but instead is acquired externally through mergers,
infalling gas clouds or through tidal stripping during close
encounters. The hot gas would then slowly flow toward the center and
settle into a cooling disk \citep[e.g.][]{BrighentiDisks}. Thus, if
angular momentum is conserved during the inflow, one should be able to
observe a trend for the X-ray ellipticities to rise inward
\citep{BregmanRotating}, as the angular momentum becomes more
important. The detailed ellipticity profiles in Figure
\ref{f.epsprofiles1} do not support this prediction observationally, since
ellipticities do not systematically increase inward. Moreover,
significant rotational support would not account for the prevalence of
strong isophotal twists.


\section{Discussion}\label{s3.discussion}

\subsection{Implications for Hydrostatic Equilibrium\label{s.implhydroeq}}

Within one stellar effective radius, the gravitational potentials of
normal elliptical galaxies are dominated by the stellar component. 
Although some studies have found dark matter contributions in individual 
objects as high as $40\%$ within this radius \citep{GerhardDarkmatter, 
PadmanabhanDarkmatter}, in general the dark matter halo contribution 
should be secondary \citep[e.g.][]{MamonDarkmatterI}. Thus, if the 
assumption of hydrostatic equilibrium were valid, one would expect the hot
gas isophotes to reflect the shape of the stellar potential
\citep{BuoteGeomtest}. Since projected isopotentials are rounder than
the projected isopleths of the underlying density, we would also
expect the gas emission to be rounder than the starlight. 
However, Figure \ref{f.epsilon} shows that the gas is flatter than
the starlight in one-third of the sample, and that gas and optical
ellipticities are completely uncorrelated. In fact, the Spearman rank-order
correlation coefficient is slightly negative, with a value of $-0.191$.
A two-sided confidence test shows that the difference from zero is not
statistically significant.\footnote{This statement also holds for
extraction annuli ranging between $0.5$ and $1.2\,R_{\rm J}$, as well as
for the sample with the disturbed objects removed.}


To determine whether the gas may still be hydrostatic despite
being morphologically uncorrelated with the starlight, we simulate
hydrostatic systems in a variety of stellar-plus-dark-matter potentials.
The stellar component is represented by a triaxial Hernquist model, whose
mass density is given by
\begin{equation}
\rho_\ast = {\rho_H \over (a/a_\ast)(1+a/a_\ast)^3},
\end{equation}
where $a$ is the semimajor axis of the ellipsoidal equidensity surface,
$a_\ast$ is the scale length, and $\rho_H$ is a normalization constant.
The dark halo is taken to be either a triaxial NFW model, having density
\begin{equation}
\label{e.nfw}
\rho_{\rm DM} = {\rho_{\rm NFW} \over (a/a_{\rm DM})(1+a/a_{\rm DM})^2},
\end{equation}
or a similar profile with a flat core:
\begin{equation}
\label{e.notnfw}
\rho_{\rm DM} = {\rho^\prime_{\rm NFW} \over (1+a/a_{\rm DM})^3}.
\end{equation}
We parametrize the models by the short-to-long
axis ratio $c/a$ and triaxiality $T = (a^2-b^2)/(a^2-c^2)$ of the stellar
and dark matter components, the ratio of scale lengths ${\bar a} \equiv
a_{\rm DM}/a_\ast$, and the ratio ${\bar M}$ of dark to luminous mass
inside the half-mass radius of the stellar component.\footnote{The half-mass
radius of the Hernquist model is $\left(1 + 2^{1/2}\right) a_\ast$, the
semimajor axis of the isodensity surface containing half the mass.}
The potentials are
computed using the formulae in the appendix of \citet{Flores}.

The principal axes of the two components
are assumed to be aligned and not intrinsically twisted.
We do not consider misaligned or twisted models because there are no
self-consistent equilibrium models for such configurations. Intrinsically
misaligned or twisted galaxies, if they exist, would likely be in some
normal mode of oscillation whose period is of order the dynamical time, and
so any gas in these systems would not be hydrostatic. We also do not
consider stellar distributions with radially varying axis ratios. Even
though this would tend to slightly decouple the local ellipticities of
the starlight and the stellar potential, this is a moot point. Only two
objects in Figure \ref{f.epsilon} show significant ellipticity gradients,
and the results are unchanged if these are removed.

The {\it a priori\/} axis ratio distribution of the stellar components is
chosen to be approximately consistent with the observed ellipticity
distribution of normal ellipticals. For populations of purely oblate or purely
prolate systems, we use rough analytic fits to the distributions obtained by
\citet{VincentRyden} from the Sloan Digital Sky Survey Data Release 3.
It is well known \citep{Ryden,TremblayMerritt}
that purely axisymmetric populations cannot reproduce
the observed ellipticity distribution without invoking negative numbers of
galaxies. We force our fits to be strictly positive, even though
this means that the observed ellipticity distribution is not exactly
recovered. For triaxial populations we use an analytic fit to the ``maximal
ignorance'' distribution of \citet{BakStatler}. This distribution is
derived from simultaneous fits to surface photometry and stellar kinematics
for 13 galaxies, and is consistent with the overall ellipticity distribution.

The combined potential is filled with isothermal gas, at a temperature
that mimics the typical observed surface brightness profiles. We orient the
system at random, and project the X-ray emissivity, which we take to be
proportional to the square of the gas density. We calculate the stellar and
gas isophotal ellipticities at $0.75$ of the stellar effective
radius,\footnote{The effective radius is the semimajor axis of the isophote
enclosing half the projected light, which depends on the axis ratios and
viewing geometry.}\ to correspond with
Figure \ref{f.epsilon}. The results are quite insensitive to the
assumed temperature; we are therefore confident that they will apply to
moderately non-isothermal equations of state as well.

\begin{figure*}
\plotone{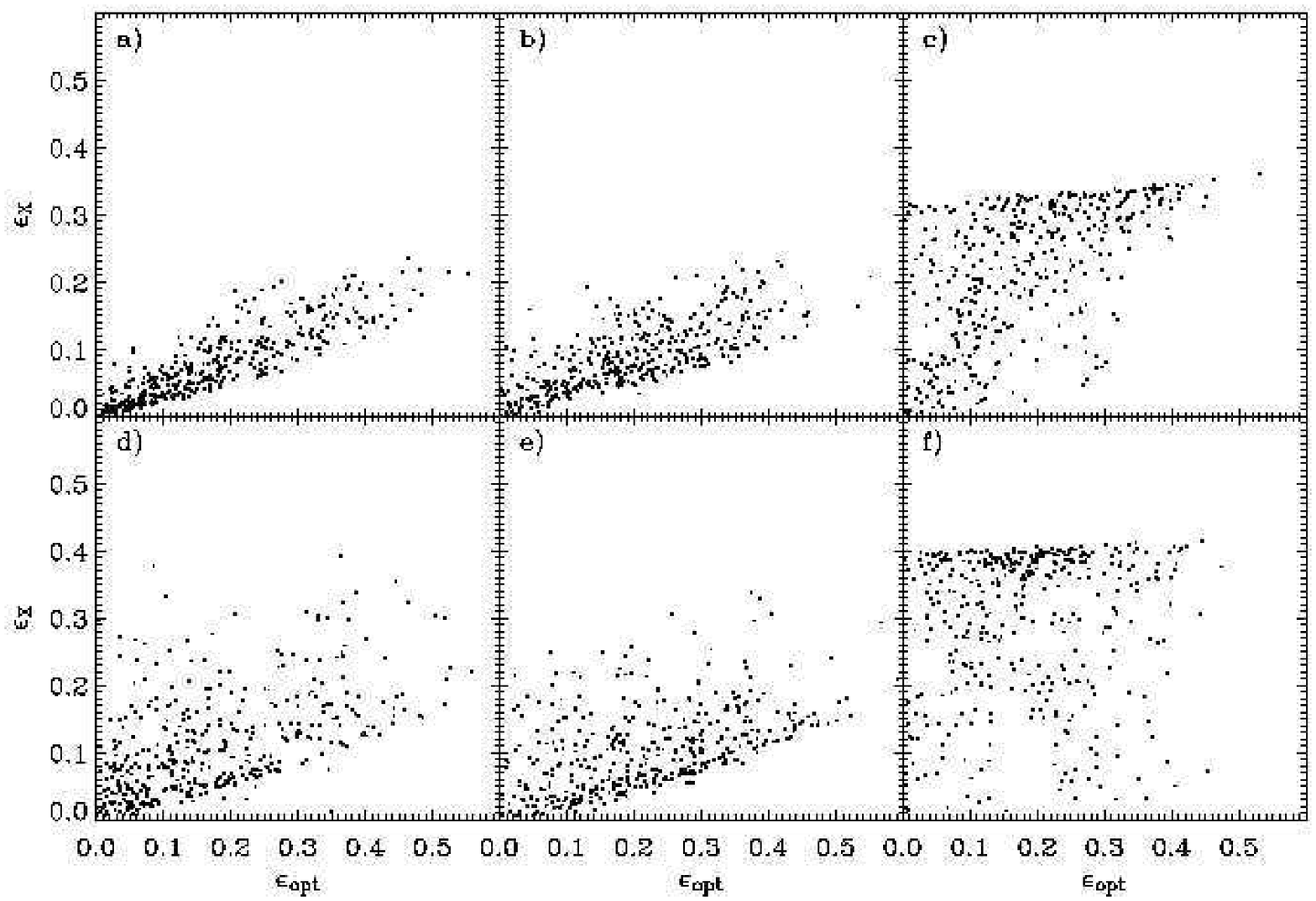}
\caption{Examples of predicted intrinsic (i.e., not convolved with measurement
errors) distributions in the $\epsilon_X$ - $\epsilon_{\rm opt}$ plane for
isothermal gas in model stellar-plus-dark-matter potentials. ($a$,$b$)
Triaxial stellar components with ($a$) strongly and ($b$) weakly correlated
halos (see \S\ 4.1); ($c$) triaxial stellar components with oblate halos
of flattening $c/a = 0.2$; ($d$) oblate and ($e$) prolate stellar components
with random triaxial halos; ($f$) triaxial stellar components with prolate
halos of flattening $c/a = 0.2$. Examples shown here all have equal contributions of dark and stellar matter (${\bar M}=1$)
and ${\bar a}=3$. Panels (a), (b), (e), and (f) are computed for NFW
halos (equation \protect{\ref{e.nfw}}); panels (c) and (d) use the flat-core
halos (equation \protect{\ref{e.notnfw}}).
\label{f.tss_simulations}
}
\end{figure*}

We first compute models in which the shapes of the dark halos are correlated
with the shapes of the stellar components. The latter are taken to be triaxial,
drawn from the \citet{BakStatler} distribution. For each object, the
halo triaxiality $T$ and axis ratio $c/a$ are chosen randomly from a gaussian
distribution centered on the stellar values, with a width $\sigma$. We
consider strongly ($\sigma = 0.2$) and weakly ($\sigma = 0.4$) correlated
halos. We compute an ensemble of several hundred models, oriented at random,
to define the predicted distribution
in the $\epsilon_X$ - $\epsilon_{\rm opt}$ plane. Examples for strongly
and weakly correlated halos are shown in Figure \ref{f.tss_simulations}a and
\ref{f.tss_simulations}b. Not surprisingly, these models predict a significant
correlation between $\epsilon_X$ and $\epsilon_{\rm opt}$.

To assess quantitatively whether the models can be consistent with the
observed distribution, we create several thousand simulated
24-object samples. For each simulated sample, we randomly perturb the
$\epsilon_X$ and $\epsilon_{\rm opt}$ values according to the actual
measurement errors in the observed sample. We then compute the Spearman
rank-order correlation coefficient and the unweighted mean ellipticity
$\langle\epsilon_X\rangle$, and determine the joint probability that the
model will produce a Spearman coefficient as low as, and a mean ellipticity
as high as, the observed values.
In assessing the models we consider ${\bar M}$ values of $0.1$, $0.3$, 1, 3,
and 10, and ${\bar a}$ values of 1, 3, and 10.

We find that all of the correlated halo models are ruled out at $>99.9\%$
confidence. This is the case even if dark mass exceeds stellar mass by as
much as a factor of 10 inside the stellar half-mass radius. Clearly, if the
isophotal ellipticity of hydrostatic gas is to be uncorrelated with that
of the starlight, the shape of the halo must be uncorrelated with that of
the stars.

Accordingly, we next consider a variety of uncorrelated halo models. These
come in two types: (1) triaxial stellar components with either purely oblate
or purely prolate halos having flattenings $c/a$ of $0.5$ and $0.2$; and (2)
axisymmetric
(oblate or prolate) stellar components with random triaxial halos. Examples
are shown in Figure \ref{f.tss_simulations}c-f. Here we
find that all of the following are excluded at $\geq 99\%$ confidence:
(1) oblate halos; (2) random triaxial halos
with mass ratios ${\bar M} \leq 3$; (3) prolate halos with $c/a=0.5$ and
${\bar M} < 3$; and (4) prolate halos with $c/a=0.2$ and ${\bar M} < 0.3$.

This result does not rely on the presence of a large number of clearly
disturbed objects in the sample. Six of the eight most asymmetric systems are
not represented in Figure \ref{f.epsilon} because they do not have published
optical surface photometry. Nonetheless, the objects with the largest
$\epsilon_X$ values in Figure \ref{f.epsilon} are NGC 4374 and NGC 5044. Of
these, the former is clearly disturbed by its radio source, and the latter
is lopsided at large radius, suggesting an interaction with the intergalactic
medium. If we cull these two objects from the sample, the constraints are
weakened slightly but our qualitative result still holds. Correlated
halo models are still ruled out at $>99.9\%$ confidence, and random triaxial
halos with mass ratios ${\bar M} \leq 3$ are still excluded at $99\%$
confidence. Oblate halos are excluded at $97\%$ confidence, and prolate halos
with $c/a=0.5$ and ${\bar M} < 3$ are excluded at $98\%$ confidence. Prolate
halos with $c/a=0.2$ cannot be excluded for ${\bar M} \geq 0.1$.

Thus the only viable hydrostatic models are those in which the shape of the
halo has nothing to do with the shape of the luminous galaxy; moreover,
the dark halo must either dominate over luminous matter inside the stellar
half-mass radius or be prominently sausage- or cigar-chaped. Both of
these options conflict with well established results.
There is widespread agreement, from stellar population modeling coupled
with stellar dynamics or X-ray data, that normal ellipticals are not
dark-matter-dominated in the inner 1 effective radius \citep{LoewWhite,
MamonDarkmatterI,Lintott,Humphrey} The mass fractions obtained in these studies
correspond to $0.2 < {\bar M} < 1.0$, which are ruled out by our simulations
except for the case of flat, prolate halos. But numerous simulations of
galaxy mergers and structure formation \citep{BarnesHernquist,Sugerman,Meza,
Bailin,Allgood,Cox} demonstrate that halos should be neither prolate nor highly
flattened. Moreover, there is no compelling reason to expect that dark
and stellar matter distributions should have very different shapes;
both obey the same collisionless dynamics, and should end up similarly
distributed following a major event such as a merger. We are therefore forced
to the conclusion that the assumption of hydrostatic equilibrium must be
incorrect.

To be more precise, even though the gas in normal ellipticals may be, on
average, close enough to hydrostatic that reasonable radial mass profiles can
be derived, hydrostatic equilibrium does not hold tightly enough that
X-ray morphology can be used to trace the shape of the total mass
distribution. Buote and collaborators \citep{BuoteNGC720,
BuoteIsotwists, BuoteGeomtest} have used this technique to argue for a
flattened dark matter halo in NGC~720. While it is possible that
hydrostatic equilibrium could hold in some individual systems, our
results show that this would be the exception rather than the rule,
and there is no independent test to reveal where hydrostatic
equilibrium does hold. Also, in the case of NGC~720, we find that properly
removing the resolved and unresolved LMXB emission reduces
the gas ellipticity, and, along with it, the need for a flattened dark
halo. The $\epsilon_X$ profile we show in Figure \ref{f.epsprofiles1}
appears to be
entirely consistent with the oblate mass-follows-light model of
\citet[Fig.~3]{BuoteNGC720} at radii between $10\arcsec$ and $80\arcsec$.
However, a direct comparison is complicated by the fact that
\citet{BuoteNGC720} do not measure {\em isophotal\/} ellipticities and
position angles, but rather their mean values {\em inside\/} elliptical
apertures.\footnote{The mass-follows-light model does not reproduce the offset
between optical and X-ray position angles, a failing that does not have a
unique interpretation.}

To preserve hydrostatic equilibrium in the general population of ellipticals,
our data would require extremely flat, complex dark matter halos with
significant substructure to explain the large ellipticities and isophotal
twists. This is highly unlikely and inconsistent with stellar
kinematics \citep[e.g.][]{SAURONCapp} or simulations of large-scale
structure formation \citep[e.g.][]{SpringelLargescalestruc}. Even
though triaxial dark matter halos are able to produce simple twists in
X-ray isophotes \citep{RomanowskyXraytwists}, these effects would be
insufficient to explain the complexity and magnitude of the observed
features.

We see no delineation between disturbed systems that are not hydrostatic
and undisturbed systems that are. While we cannot prove that the round,
symmetric systems are not hydrostatic, we emphasize that the sample, as a
whole, has failed the test. If one chooses to invoke hydrostatic equilibrium
for a system that appears ``relaxed,'' it should be clearly understood that
this reflects a hope rather than a verifiable assumption.

We take the view that the isophotal ellipticity of the X-ray gas does not
reflect the underlying potential. Ellipticity is merely a measure of the
$m=2$ Fourier amplitude of the deviations from circular symmetry. In the
optical, the other Fourier harmonics are small by comparison; in the X-ray,
they are not. In Paper II of this series we will argue that the gas
ellipticity is a measure of morphological disturbance, quantify that
disturbance, and demonstrate that it is likely produced by AGN.

The assumption of hydrostatic equilibrium is 
widely used to derive radial mass profiles
\citep[e.g.][]{Forman1985,KilleenEinsteindarkmatter,FabbianoNGC507,
HumphreyDarkmatter,FukazawaMassprofiles}. If the
gas is not hydrostatic, what are the consequences?
Imagine a region that was locally overpressured by a factor
$q$. Assuming adiabatic expansion this region would expand by a factor
$q^{1/5}$ in linear size over a sound crossing time scale of $\sim
10^8$ years, to regain pressure equilibrium. To redistribute gas a
significant distance along an equipotential would thus imply an
overpressure $q \lesssim 10$. If the galaxy were {\em globally\/}
overpressured by this factor, one would infer a mass a factor of $q$
too large. However, normal ellipticals are probably only locally
overpressured, and so averaging azimuthally over $N$ over- and
underpressured regions in a radial analysis would tend to reduce this
effect by a factor of order $N^{1/2}$, resulting in a mass estimate in
error by at most a factor of a few, either high or low. Subsonic
bulk motions generated by such temporary overpressures could also enhance or
reduce radial pressure support, depending on their radial gradient.

An early study by \citet{SchindlerMassestimates} demonstrated that mass
profiles of simulated clusters can be recovered successfully with the
hydrostatic equilibrium method, limited primarily by the presence of significant
substructure. More recent studies of hydrodynamic cluster formation
simulations \citep{Rasia2004,Nagai2007} show that neglecting bulk motion 
in a hydrostatic analysis tends to underestimate the gravitating mass by
5 to 20\%.
\citet{RasiaMassestimates} find a further systematic effect of similar
magnitude that they attribute to the complex temperature structure of the
cluster; however this appears only when they reduce the backgrounds in
their simulated observations far below typical {\em Chandra\/} values.
Unfortunately, no such study is available for the mass regime of normal
elliptical galaxies and none yet takes into account disturbances due to
AGN activity, which our simple arguments above suggest may be larger in
magnitude. 
On the observational side, \citet{FukazawaMassprofiles} derive total mass
profiles from {\it
Chandra} X-ray data of 53 normal elliptical galaxies assuming hydrostatic
equilibrium and compare 7 of these to mass profiles based on stellar
kinematics. They find that for 2 out of 7 galaxies, total enclosed masses
(inside observationally accessible maximum radii)
differ by a factor of $\sim 2$ and for one object,
NGC~3379, the difference even grows to a factor of $\sim
7$. \citet{PellegriniNGC3379Wind} reconcile this discrepancy for
NGC~3379 by fitting its X-ray properties with a wind model, staying
consistent with the optically derived mass profile. They conclude
that the gas is not in hydrostatic equilibrium, but in a general
state of outflow, calling the X-ray mass measurement into question.

\subsection{Implications for Rotational Support}

The hot gas in elliptical galaxies is believed to come from some
combination of stellar mass loss, infall or mergers. In each case, the
gas should carry significant amounts of angular momentum. In a
standard cooling flow model that conserves angular momentum, the gas
will settle into rotationally supported cooling disks
\citep{BrighentiDisks,BrighentiDisks2,KleyXraydisks}. These rotating
cooling flow models predict ellipticity profiles that rise inward,
with minimal isophotal twist.

Figures \ref{f.epsprofiles1} and \ref{f.vroteps} show that these
predictions are inconsistent with observations. Although we do find
rather large X-ray ellipticities, they do not systematically increase
toward smaller radii. Our profiles also reveal dramatic isophotal
twists and asymmetries that would not arise naturally in a scenario
where the ellipticities are caused by rotation alone. In addition, we
find no relation between stellar rotational velocity and the gas
flattening, as one would expect if the gas is mainly due to stellar
mass loss. Thus, we conclude that the X-ray gas morphology is not
dictated by rotation, in agreement with previous ROSAT observations
\citep{BregmanRotating}.

The question now is whether the failure to observe rotationally
flattened X-ray disks is a serious blow to cooling flow models. One
way to save the cooling flow picture would be to efficiently transfer
angular momentum through turbulence in the gas
\citep{ShadmehriRotatingviscous,BrighentiRotating}, but it is unclear
how effective this process is for elliptical galaxies. Another
alternative is that spatially distributed multi-temperature mass
dropout can circularize the X-ray isophotes
\citep{BregmanRotating}. However, recent XMM-Newton spectroscopy of
normal elliptical galaxies \citep[e.g][]{XuNGC4636XMM,
BuoteNGC5044XMM} rules out the existence of sufficient intermediate
temperature gas. In addition, \citet{BregmanExtendedOVI} are able to
directly trace the amount of this intermediate gas by measuring its O
VI emission ($\sim 10^{5.5}{\rm K}$) with the {\it Far Ultraviolet
Spectral Explorer} (FUSE). \citet{BregmanOVI} examine the spatial
distribution of this warm gas for NGC~4636 and NGC~5846 and 
constrain
the size of the cooling region to be smaller than $0.8\kpc$ and
$0.5\kpc$, respectively. Their measurements are consistent with
moderate cooling flows with centrally concentrated mass dropout and
rule out spatially distributed dropout for these galaxies.

Another explanation for the lack of disk signatures is that the hot
gas is actually flowing outward instead of inward, eliminating the
need for the gas to settle into a cooling disk. An idea by
\citet{BrighentiMassiveoutflows} seems to be successful in stopping
inward cooling flows by AGN-induced massive jet outflows, which are
stable for several gigayears.

Alternatively, morphological asymmetries, which are clearly
significant, may be masking possible disk signatures. Whatever the
cause of these asymmetries (see Paper II), this effect could be
effective in disrupting the flattest X-ray isophotes at small radii
close to the center. Detailed simulations are needed to confirm this
possibility. Thus, our results do not imply that rotational support of
the hot gas in elliptical galaxies has no importance at all. We can
merely state that it is not the {\em dominant} factor that causes the
large observed ellipticities.

There is other evidence that gas rotation might be present. For
example, \citet{NGC1700} find extremely flat X-ray emission in
NGC~1700, which they interpret as a large-scale cooling disk. Their
model yields a specific angular momentum and cooling time for the hot
gas that is consistent with the gas having been acquired during the
last major merger. We find that removal of unresolved point sources
somewhat reduces the X-ray ellipticity of this object; but the data
are still consistent with their claimed $15\kpc$ rotating
disk. Additional support for rotation has recently been reported by
\citet{BregmanOVI}, who find that the O VI line structure of NGC~4636
exhibits signs consistent with rotation.

\section{Conclusions}\label{s3.conclusions}

We have analyzed the X-ray emission of 54 normal elliptical galaxies
in the {\it Chandra} archive and isolated their diffuse hot
interstellar gas emission from the emission of discrete stellar point
sources for the first time. We qualitatively and quantitatively
compare the morphology of the hot gas to the shape of the stellar
distribution and find that they have very little in common, despite
the known $L_{\rm X}$--$L_{\rm B}$ relation. We compute ellipticity
and position angle profiles for the X-ray gas and compare them to
published optical profiles. In particular, we do not find a
correlation between optical and X-ray ellipticities measured in the regions
where stellar mass dominates the potential, suggesting that
the gas is at least far enough out of hydrostatic equilibrium that the
information about the shape of the underlying potential is lost. We
also argue that X-ray derived radial mass profiles may be in error by
factors of as much as a few, without the necessity for the galaxy to
drive a global wind.

Although we find large X-ray gas ellipticities to be common, the
gas morphology is generally inconsistent with rotationally flattened
disks, and a comparison with stellar rotational velocities yields no
evidence for significant rotational support.

The fact that neither the shape of the underlying potential nor
rotational support determine the overall distribution of the X-ray
emitting gas, combined with its general disturbed appearance, suggests
the involvement of another major component: the central AGN. We 
assess
the importance of the AGN in Paper II, where we draw a connection
between gas morphology and AGN luminosity. These new findings are
consistent with the AGN constantly stirring up the interstellar medium
by inflating buoyant bubbles, which may also play a role in
redistributing the angular momentum of the hot gas through
entrainment. These intermittent AGN outbursts could also be
responsible for disrupting or masking the signatures of cooling disks
in the central regions of elliptical galaxies.


\acknowledgments We would like to thank Jesus Falc\'on-Barroso and 
the
SAURON team for allowing us to use their yet unpublished MDM
photometry. We have made use of the HyperLEDA database
(http://leda.univ-lyon1.fr) and of data products from the Two Micron
All Sky Survey, which is a joint project of the University of
Massachusetts and the Infrared Processing and Analysis
Center/California Institute of Technology, funded by the National
Aeronautics and Space Administration and the National Science
Foundation.

We also made use of the Digitized Sky Surveys which were produced at
the Space Telescope Science Institute under U.S. Government grant 
NAG
W-2166. The ``Second Epoch Survey'' of the southern sky was produced
by the Anglo-Australian Observatory (AAO) using the UK Schmidt
Telescope. The digitized images are copyright (c) 1993-1995 by the
Anglo-Australian Telescope Board, and are distributed herein by
agreement. The ``Equatorial Red Atlas'' of the southern sky was
produced using the UK Schmidt Telescope. Plates from both surveys 
have
been digitized and compressed by the ST ScI. The digitized images are
copyright (c) 1992-1995, jointly by the UK SERC/PPARC (Particle
Physics and Astronomy Research Council, formerly Science and
Engineering Research Council) and the Anglo-Australian Telescope
Board, and are distributed herein by agreement. All Rights
Reserved. The compressed files of the ``Palomar Observatory - Space
Telescope Science Institute Digital Sky Survey'' of the northern sky,
based on scans of theSecond Palomar Sky Survey, are copyright (c)
1993-1995 by the California Institute of Technology and are
distributed herein by agreement. All Rights Reserved.

Support for this work was provided by the National Aeronautics and 
Space
Administration (NASA) through Chandra Awards G01-2094X and 
AR3-4011X,
issued by the {\em Chandra X-Ray Observatory Center}, which is 
operated by the
Smithsonian Astrophysical Observatory for and on behalf of NASA under
contract NAS8-39073, and by National Science Foundation grant 
AST0407152. Any opinions, findings, and conclusions or recommendations
expressed in this material are those of the authors and do not necessarily
reflect the views of the National Science Foundation.

\bibliographystyle{apj}
\bibliography{/Users/diehl/papers/bibtex/allreferences.bib}

\end{document}